\documentclass[preprint,12pt]{elsarticle}

\usepackage{amssymb}
\usepackage[dvipsnames]{xcolor}
\usepackage{color}
\usepackage[version=3]{mhchem} % Formula subscripts using \ce{}

\usepackage{times}
\usepackage{amsmath}
\usepackage{amssymb}
\usepackage{bm}
\usepackage{here}
\usepackage{graphicx}
\usepackage{framed}
\definecolor{shadecolor}{rgb}{0.9,0.9,0.9}
\usepackage[utf8x]{inputenc}

\journal{hhh}

\begin{document}

\begin{frontmatter}



\title{Superconductivity in CH$_4$ and BH$_4^-$ Containing Compounds Derived from the High-Pressure Superhydrides}


\author[label1]{Nisha Geng\fnref{myfootnote}}

\affiliation[label1]{organization={Department of Chemistry},%Department and Organization
            addressline={ State University of New York at Buffalo}, 
            city={Buffalo},
            postcode={14226-3000}, 
            state={NY},
            country={USA}}

\author[label1]{Katerina P. Hilleke\fnref{myfootnote}}
\author[label1]{Francesco Belli}
\author[label1]{Pratik Kumar Das}
\author[label1]{Eva Zurek\corref{mycorrespondingauthor}}\cortext[mycorrespondingauthor]{Corresponding author}
\ead{ezurek@buffalo.edu}
\fntext[myfootnote]{These authors contributed equally}

\begin{abstract}
Inspired by the synthesis of the high-pressure $Fm\bar{3}m$ LaH$_{10}$ superconducting superhydride, systematic density functional theory (DFT) calculations are performed to study ternaries that could be derived from it by replacing two of the hydrogen atoms with boron or carbon and varying the identity of the electropositive element. Though many of the resulting alkali-metal and alkaline-earth $M$C$_2$H$_8$ phases are predicted to be dynamically stable at mild pressures, their superconducting critical temperatures ($T_c$s) are low because their metallicity results from the filling of an electride-like band. Substitution with a trivalent element leads to phases with substantial metal $d$-character at the Fermi level whose $T_c$s are typically above 40~K. Among the $M$B$_2$H$_8$ phases examined, KB$_2$H$_8$, RbB$_2$H$_8$ and CsB$_2$H$_8$ are predicted to be dynamically stable at very mild pressures, and their stability is rationalized by a DFT-Chemical Pressure analysis that elucidates the role of the $M$ atom size. Quantum anharmonic effects strongly affect the properties of KB$_2$H$_8$, the highest predicted $T_c$ compound, near 10~GPa, but molecular dynamics simulations reveal it would decompose below its $T_c$ at this pressure. Nonetheless, at ca.\ 50~GPa KB$_2$H$_8$ is predicted to be thermally stable with a superconducting figure of merit surpassing that of the recently synthesized LaBeH$_8$.
\end{abstract}




\end{frontmatter}


\newpage

\section{Introduction}
The search for hydrogen-rich materials with high superconducting critical temperatures ($T_c$s) has made great strides~\cite{Zurek:2020k} in the two decades since Neil Ashcroft predicted that doping hydrogen with a second element could expedite its metallization thereby inducing superconductivity~\cite{Ashcroft:2004a}. Ashcroft argued that the internal pressure applied by the doping atoms would ``chemically precompress" the main hydrogen matrix, and lower the physical pressure required for metallization. Since then, a number of chemically intriguing and unexpected hydride phases with high $T_c$s have been successfully synthesized at pressures attainable in diamond anvil cells including H$_3$S ($T_c=$~203~K at 150~GPa~\cite{Drozdov:2015}), LaH$_{10}$ ($T_c=$~260~K at 200~GPa~\cite{Somayazulu:2019,Drozdov:2019}), YH$_6$ ($T_c=$~220~K at 183~GPa~\cite{Kong:2021a}, or 224~K at 166~GPa~\cite{Troyan-YH4}), CaH$_6$ ($T_c=$~215~K at 172~GPa~\cite{Ma:CaH6}, or 210~K at 160~GPa~\cite{Li:CaH6}), (La,Y)H$_6$, ($T_c=$~237~K at 183~GPa~\cite{Semenok:2021}), and YH$_9$ ($T_c=$~262~K at 182~GPa~\cite{Kong:2021a}). However, challenges still remain: while these materials exhibit impressive $T_c$s, all require substantial pressure to prevent their decomposition. In fact, the discovery of these superconducting hydrides has spawned a new quest, this time for hydrogen-based high-$T_c$ phases that remain stable at or near atmospheric pressures~\cite{Zurek:2021k}. 

The  pursuit of high-$T_c$ superconductors has been greatly aided by the increasing speed and power of first-principles calculations. Since high-pressure experiments are expensive and difficult, much cheaper computations can screen candidate systems and identify promising leads. Crystal structure prediction (CSP) techniques~\cite{Zurek:2018m,Flores-Livas:2020} have been key in this aspect. In fact, several of the previously mentioned high-$T_c$ materials were first uncovered via CSP studies~\cite{Wang:2012, Liu:2017, Peng:2017} prior to their experimental realization. These are the ``clathrate-like" superhydrides, including those with $M$H$_6$, $M$H$_9$, and $M$H$_{10}$ stoichiometries in which weakly covalently-bonded hydrogen atoms form three-dimensional networks that are able to encapsulate metal atoms. In contrast, $Im\bar{3}m$ H$_3$S belongs to another family of hydride superconductors built with covalent bonds between hydrogen and a main group non-metal~\cite{Drozdov:2015,Duan:2014}. The binary $M$-H phase diagrams have been fairly exhaustively explored, with plenty of promise but still, no room-temperature ambient-pressure superconductors have been discovered~\cite{Zurek:2018d}. Therefore, researchers have begun searching for superconductivity in ternary and even quaternary hydride phases. 

\begin{figure*}
\begin{center}
\includegraphics[width=0.8\columnwidth]{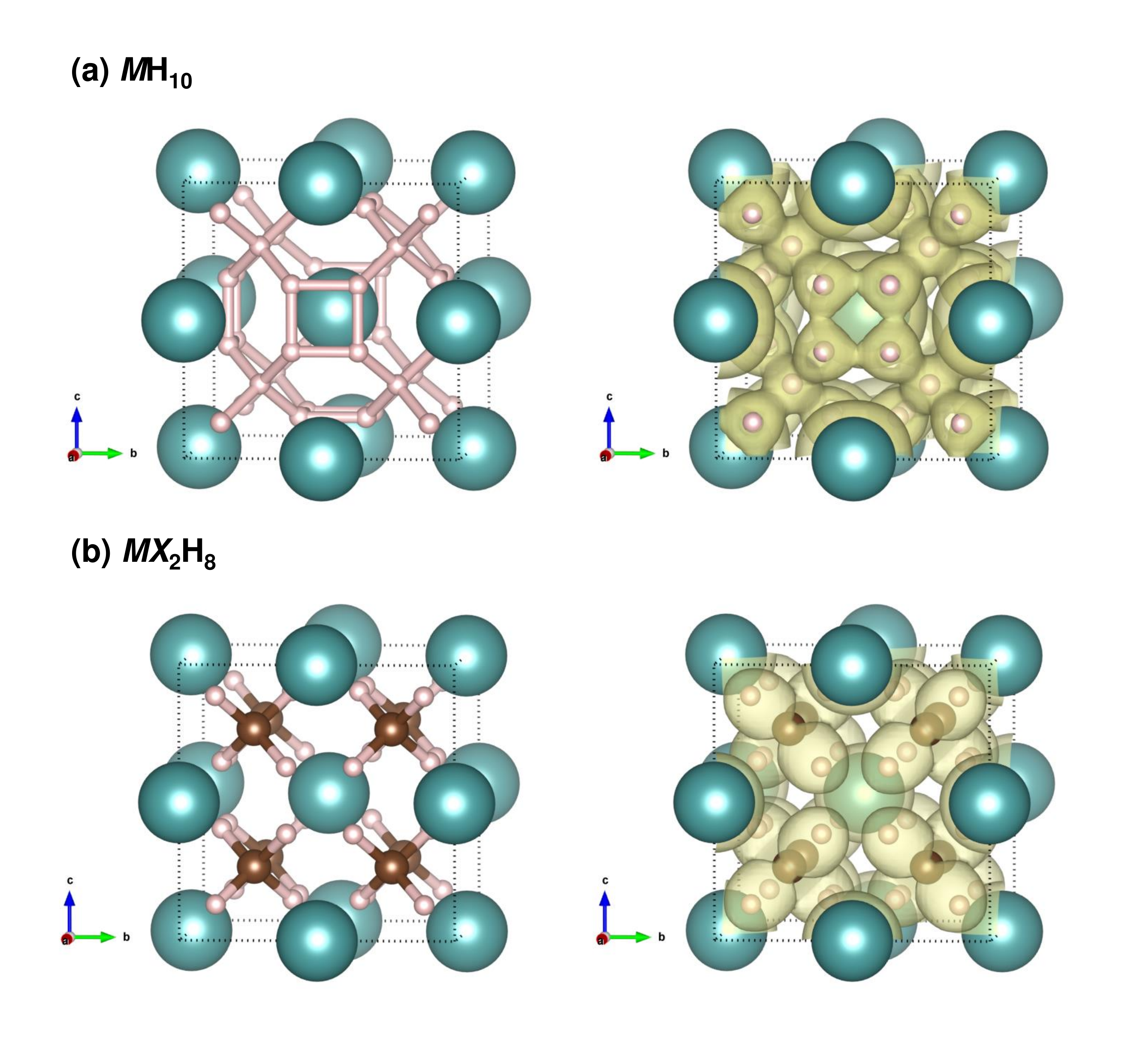}
\caption{Illustrations of structure (left) and electron localization function plot (right) of the conventional cell of: (a) $Fm\bar{3}m$ $M$H$_{10}$ and (b) $Fm\bar{3}m$ $MX$$_2$H$_8$, where $M$ is an electropositive metal atom and $X$ denotes carbon or boron. $M$/H/$X$ atoms are colored green/pink/brown. Isosurface = 0.5 computed for $Fm\bar{3}m$ LaH$_{10}$ and $Fm\bar{3}m$ KB$_2$H$_8$ at 300~GPa.}
    \label{fig:mx10structure}
\end{center}
\end{figure*}

A major obstacle, however, is the vast number of combinatorial possibilities that render Density Functional Theory (DFT) based CSP searches on a broad range of chemical combinations unfeasible. One way to lower the tremendous computational cost is via the generation of machine-learned interatomic potentials that can roughly scan the phase space for promising compositions~\cite{Salzbrenner:2023, Ferreira:2023,Pickard:2023f}, which are later more thoroughly explored using DFT. In many cases, however,  more traditional methods have come to the fore in the prediction of novel hydrides. One approach is to combine two already known superconducting binaries. For instance, a study that identified a $P\overline{6}m2$ CaSH$_3$ phase with a computed $T_c$ of $\sim$100~K at 128 GPa~\cite{Yan:2020} was inspired by the prediction and synthesis of CaH$_6$~\cite{Wang:2012, Ma:CaH6, Li:CaH6} and H$_3$S~\cite{Drozdov:2015,Duan:2014}.  A second approach is to add other elements into previously identified superconducting compounds, exemplified by substitutional doping in H$_3$S-based materials \cite{Geng:2022,Ge:2020a}. Finally, DFT-based high-throughput searches on prototype structures that have been discovered either via CSP~\cite{Zurek:2023a}, or by ternary modifications of the binary superhydrides~\cite{Lucrezi:2022} have identified promising candidates that present a balance between high-$T_c$ and low stabilization pressure. For example, theoretical predictions have suggested that two main modifications of $Fm\bar{3}m$ $M$H$_{10}$ (Figure\ \ref{fig:mx10structure}a)  -- the structure assumed by the first clathrate superhydride to be synthesized, LaH$_{10}$ -- result in ternary hydrides that can be quenched to lower pressures than the binary. The hydrogenic lattice in this structure can be viewed as a network of vertex-sharing H$_4$ tetrahedra and H$_8$ cubes. One pathway, by removing a hydrogen atom from the  8$c$ Wyckoff site that lies in the center of the tetrahedra, and stuffing the H$_8$ cubes with Be or a $p$-block element results in phases with $MX$H$_8$ compositions \cite{Zhang:2021,Lucrezi:2022,Liang:2021a,Di:2021bh},  exemplified by LaBeH$_8$, which was synthesized and measured to have a $T_c$ up to 110~K at 80~GPa~\cite{Song:2023s}.

The focus of the present investigation is a different modification of $Fm\bar{3}m$ $M$H$_{10}$, attained by replacing the hydrogen atoms at the 8$c$ site with $X$ (here, B or C) yielding the $MX_2$H$_8$ structure (Figure\ \ref{fig:mx10structure}b). The pressure-dependence of the dynamic stability and superconducting properties of a few $MX_2$H$_8$ compounds has already been explored computationally including $M$C$_2$H$_8$ \cite{Jiang:2022,Zurek:2022f,Durajski:2021}, $M$B$_2$H$_8$ \cite{Li:2022s,Gao:2021}, and $M$N$_2$H$_8$ \cite{Wan:2022}. These initial studies suggest this family of compounds may present a viable route towards warm and light superconductivity manifested by KB$_2$H$_8$, which has been predicted to maintain dynamical stability as low as 12~GPa ($T_c$ = 146~K)~\cite{Gao:2021}, LaC$_2$H$_8$ with $T_c=$~69~K at 70~GPa~\cite{Durajski:2021}, and YC$_2$H$_8$ with $T_c$ = 61~K at 50~GPa~\cite{Zurek:2022f}. The latter of these compounds was constructed on the basis of DFT-Chemical Pressure arguments, which interrogate the internal stresses within a crystal structure that arise due to atomic size. Unfortunately, since $T_c$ estimates can be extremely sensitive to the methodology and the computational settings used, it is difficult to uncover trends in the aforementioned studies,  which typically looked at no more than a few compositions. 

Therefore, herein we performed systematic high-throughput calculations on $Fm\bar{3}m$ $MX_2$H$_8$ compounds varying the identity of the electropositive metal atom ($M$= Li, Na, K, Rb, Cs, Be, Mg, Ca, Sr, Ba, Sc, Y, and La) for both B and C substitutions. Their dynamic stability, electronic structure and propensity for superconductivity was computed, as a function of pressure from ambient to 300~GPa. Such studies can uncover a plethora of chemical trends, for example the connection between stability and ionic radii of the electropositive element, and oxidation state of the metal atom with the predicted $T_c$, facilitating the comparison of the properties of a single structure with different elemental compositions~\cite{Geng:2023}. We find that from the 26 different $M$B$_2$H$_8$ and  $M$C$_2$H$_8$ compounds considered, 14 are (phonon) metastable at various points from ambient to 300~GPa, some to as low as 4~GPa. Though a number of the $M$C$_2$H$_8$ species were metallic and stable, only those containing a trivalent metal element had appreciable $T_c$s below $\sim$200~K. In addition, we considered the effects of atomic size on dynamical stability and the role of valence electron count on $T_c$, finding that univalent $M$ atoms yield both a strong electron phonon coupling and high $T_c$ even at low pressures within $M$B$_2$H$_8$ compounds. Moreover, quantum nuclear effects were found to renormalize the frequencies of the phonon modes and resulting electron-phonon coupling strength within KB$_2$H$_8$, the phase with the highest predicted $T_c$ at the lowest pressure, which was predicted to thermally decompose below its $T_c$ at 12~GPa. Molecular dynamic simulations suggest that at least 50~GPa is required to stabilize KB$_2$H$_8$ to temperatures surpassing its isotropic Eliashberg $T_c$, 85~K.

\section{Computational Details}
Geometry optimizations, molecular dynamics simulations, and electronic structure calculations including band structures, densities of states (DOS), electron localization functions (ELF), and Bader charges were calculated by density functional theory (DFT) as implemented in the Vienna \textit{ab-initio} Simulation Package (VASP) version 5.4.1 \cite{Kresse:1993a, Kresse:1999a}. The gradient-corrected exchange and correlation functional of Perdew{-}Burke{-}Ernzerhof (PBE) \cite{Perdew:1996a} was employed along with the projector augmented wave (PAW) method \cite{Blochl:1994a}. The plane-wave basis set energy cutoff was 900~eV and the B 2s$^2$2p$^1$, C 2s$^2$2p$^2$, and H 1s$^1$ electrons were treated explicitly in all of the calculations. The valence configurations and PAWs employed for the metal atoms are listed in Table S1. The $\Gamma$-centered Monkhorst-Pack scheme was used to produce $k$-point meshes chosen so that the product of the number of divisions along each reciprocal lattice vector with the real lattice constant was 50~\AA{} in the geometry optimizations, and 70~\AA{} otherwise. The crystal orbital Hamilton population (COHP) and the negative of the COHP integrated to the Fermi level (-iCOHP), calculated by the LOBSTER package~\cite{COHP-2016}, were used to estimate the bond strengths within selected phases.  The optimized structural parameters are provided in Table S17-S18. The thermal stability of KB$_2$H$_8$ at 3, 12, and 50~GPa (and YC$_2$H$_8$ at 50~GPa)  was examined by performing \emph{ab initio} molecular dynamics (AIMD) simulations using a canonical $NVT$ ensemble at 77, 165, and 300~K, with temperature and volume controlled via a Nos\'e-Hoover thermostat~\cite{Nose:1984,Shuichi:1991,Hoover:1985,Frenkel:2023}. A $2\times2\times2$ supercell was chosen to reduce the constraint of periodic boundary conditions with an energy cutoff of 600~eV, and only the $\Gamma$-point was used. All AIMD simulations contained 10,000~MD steps (10~ps). High-throughput phonon calculations were carried out using the VASP package coupled with the PHONOPY code~\cite{Togo:2015}. 

The electron-phonon coupling (EPC) calculations were performed with the Quantum Espresso (QE) program package~\cite{Giannozzi:2009}. The  B 2s$^2$2p$^1$, C 2s$^2$2p$^2$, H 1s$^1$ and metal pseudopotentials (Table S1) were obtained from the PSlibrary~\cite{DalCorso:2014}, and generated by the method of Trouiller-Martins~\cite{Troullier:1991}. The plane-wave basis set energy cutoffs varied from 80-120~Ry (Table S2). The Brillouin zone sampling scheme of Methfessel-Paxton \cite{Methfessel:1989} was applied with a smearing width of 0.01-0.03~Ry (Table S2). A $16\times16\times16$ $k$-point grid was used for all phonon calculations, while a dense $32\times32\times32$ $k$-point grid and an $8\times8\times8$ $q$-mesh was used for all of the EPC calculations. The EPC parameter, $\lambda$, was calculated using a set of Gaussian broadenings from 0.0 to 0.500~Ry (with an increment of 0.005~Ry) and converged to 0.05~Ry. The $T_c$s were estimated using the Allen-Dynes modified McMillan equation \cite{Allen:1975} with a renormalized Coulomb potential, $\mu^*$, of 0.1. For those systems with an EPC constant, $\lambda$, larger than unity the $T_c$s were calculated via numerical solution of the Eliashberg equations~\cite{Eliashberg:1960}. The quantum nuclear effects and anharmonicity on the dynamical stability and superconducting properties of $Fm\bar{3}m$ KB$_2$H$_{8}$ were studied using the Stochastic Self Consistent Harmonic Approximation (SSCHA)~\cite{Monacelli:2021t}. The quantum anharmonic phonon spectra was calculated at 0~K, in a $2\times2\times2$ supercell, using up to 3000 configurations containing 88 atoms each, and including terms up to the fourth order in the calculation of the free energy Hessian. The self consistent calculations required for the stochastic sampling were performed using the QE program package with parameters commensurate to those specified above. 

A DFT-Chemical Pressure Analysis~\cite{Fredrickson:2012}, which visualizes the internal stresses inherent in a crystal structure as a consequence of steric constraints, was performed on select phases using the \emph{CPpackage2}~\cite{Berns:2014a} coupled with output from relevant modules~\cite{Hilleke:2018,Lu:2021}, as described fully in Section S1.3. The compounds examined: LiB$_2$H$_8$, NaB$_2$H$_8$, KB$_2$H$_8$, RbB$_2$H$_8$, and CsB$_2$H$_8$, were optimized at 0~GPa using the ABINIT~\cite{Gonze:2020} software package and LDA-DFT (Section S8).

\section{Results and Discussion}

\begin{figure*}
\begin{center}
\includegraphics[width=1\columnwidth]{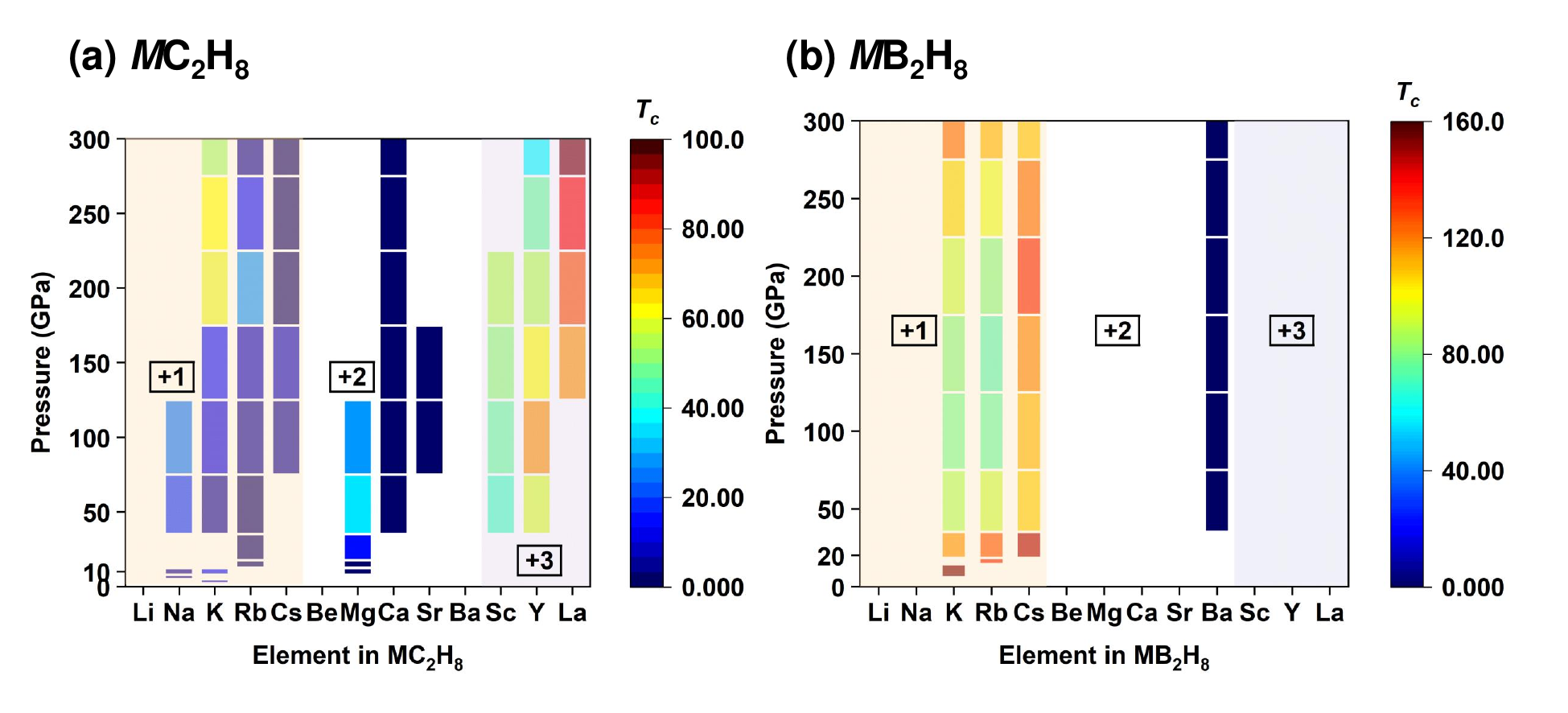}
\caption{A map of the superconducting critical temperature, $T_c$, calculated via numerical solution of the Eliashberg equations and assuming a Coulomb repulsion parameter, $\mu^*=$~0.1, of  (a) $M$C$_2$H$_8$ and (b) $M$B$_2$H$_8$ from 0~GPa to 300~GPa with 50~GPa increments. The identities of the metal atoms are given on the $x$-axes, with the various regions color coded by the metal valence (+1, +2, or +3). For the lower pressures the $T_c$ map is colored by values obtained in the middle of the pressure interval, while for pressures above 50~GPa the $T_c$ was calculated at the end of the pressure interval (Table S9-S11).}
\label{fig:mx10tc}
\end{center}
\end{figure*}

Our high-throughput DFT calculations showed that fourteen out of the twenty-six chemical compositions studied were dynamically stable, having no imaginary phonon frequencies throughout the whole Brillouin Zone, within an element-dependent pressure range. From these, seven -- NaC$_2$H$_8$ (6~GPa), KC$_2$H$_8$ (4~GPa), RbC$_2$H$_8$ (15~GPa), MgC$_2$H$_8$ (10~GPa), KB$_2$H$_8$ (12~GPa), RbB$_2$H$_8$ (16.5~GPa) and CsB$_2$H$_8$ (20~GPa) -- were dynamically stable at the near-ambient pressures that are given in the braces. The remaining seven --  CsC$_2$H$_8$, CaC$_2$H$_8$, SrC$_2$H$_8$, ScC$_2$H$_8$, YC$_2$H$_8$, LaC$_2$H$_8$, and BaB$_2$H$_8$  -- required at least 50~GPa to prevent their vibrations from inducing structural phase transitions. These results are plotted graphically in Figure\ \ref{fig:mx10tc}, where the dynamically stable phases are denoted by rectangles color-coded according to their predicted $T_c$ from blue (cold) to red (hot). Insulating compounds are denoted by a dark blue rectangle, and those that are dynamically unstable are left empty.

Some of these systems have been studied computationally before. Yttrium~\cite{Zurek:2022f} and lanthanum-containing~\cite{Durajski:2021} $Fm\bar{3}m$ $M$C$_2$H$_8$ phases were predicted to remain dynamically stable to pressures as low as 50~GPa and 70~GPa, respectively.  Jiang \emph{et al.}\ uncovered several more $M$C$_2$H$_8$ phases containing Na (20-100~GPa), K (50-100~GPa), Mg (20-40~GPa), Al (10-100~GPa), and Ga (20-100~GPa) to be dynamically stable in the pressure ranges provided in the braces~\cite{Jiang:2022}. Turning to the boron containing compounds, Gao \emph{et al.}\ and Durajski \emph{et al.}\ found $Fm\bar{3}m$ KB$_2$H$_8$ to be dynamically stable from 12-200~GPa~\cite{Gao:2021} and 3-100~GPa~\cite{Durajski:2023first}. A later CSP study by Li \emph{et al.} identified this to be the most stable  KB$_2$H$_8$ phase between 30-100~GPa, but at lower pressures an $I4_1/a$ symmetry structure was preferred instead~\cite{Li:2022s}. CSP was also employed to search for the most stable $M$B$_2$H$_8$ phases across the range of alkali metal atoms (excluding francium) and only KB$_2$H$_8$, RbB$_2$H$_8$ and CsB$_2$H$_8$ were computed to be dynamically stable above 10~GPa~\cite{Li:2022s}. In a related work, the pressure dependant dynamic stability of ammonium containing systems with the general formula $M$N$_2$H$_8$ and a variety of metal atoms was computationally investigated~\cite{Wan:2022}. As described below, our study uncovered six dynamically stable species that were not predicted in these previous investigations: RbC$_2$H$_8$, CsC$_2$H$_8$, CaC$_2$H$_8$, SrC$_2$H$_8$, ScC$_2$H$_8$ and BaB$_2$H$_8$, and extended the pressure range probed for dynamical stability from 4~GPa to 300~GPa. Let us now examine the electronic structure and propensity for superconductivity in these compounds.

In contrast to their cousin $M$H$_{10}$ and $MX$H$_8$ structures (e.g.\ LaH$_{10}$ and LaBH$_8$), for which the hydrogen atoms adopt a loose covalent 3-D network, the $MX_2$H$_8$ phases examined herein present a rather different bonding environment. The placement of the B or C atoms in the metal hydrogen framework renders them tetrahedrally coordinated by hydrogen atoms, leading to the formation of molecular  BH$_4^-$ or CH$_4$ units. Whereas the ELF isosurface encapsulates the 3-D hydrogenic framework in $M$H$_{10}$, for  $MX_2$H$_8$ high values of the ELF, suggestive of covalent bonding, are present along the B-H and C-H contacts, but the ELF between hydrogen atoms in  neighboring BH$_4^-$ or CH$_4$ fragments is very low (e.g.\ see Figure\ \ref{fig:mx10structure}).  For example, for KB$_2$H$_8$, the ELF reaches a minimum value of 0.14 at the midpoint between these hydrogen atoms (separated by 1.90~\AA{} at 12~GPa).  Increasing the pressure to 300~GPa pushes the BH$_4^-$ units closer together so that the shortest distance between their hydrogens measures 1.20~\AA{}, but the ELF at the midpoint increases only slightly to 0.37. In both cases, there is a sharp delineation between the ELF isosurfaces associated with different BH$_4^-$ subunits.  For comparison, in $Fm\bar{3}m$ LaH$_{10}$ at 150~GPa, the ELF value reaches a minimum of 0.64 between the 8$c$ and 32$f$ H atoms and 0.52 for the contacts between 32$f$ H atoms (analogous to the inter-BH$_4^-$ H-H contacts in KB$_2$H$_8$).

How might we understand the curious trends seen in Figure\ \ref{fig:mx10tc}, as to which metal atoms result in dynamically stable $MX_2$H$_8$ phases? One difficulty arises from the fact that the $M$B$_2$H$_8$ and $M$C$_2$H$_8$ systems appear to require different metal atoms to be phonon-stable (at some pressure in the range considered): for $M$C$_2$H$_8$ this includes the alkaline earths (except Be and Ba) as well as trivalent Sc, Y, and La. In $M$B$_2$H$_8$ phases, on the other hand, dynamical stability is only achieved with barium (from the alkaline earths) and none of the trivalent phases correspond to local minima. The only commonality is in a subset of the alkali metals: K, Rb, and Cs. Thus, for both $M$B$_2$H$_8$ and $M$C$_2$H$_8$ systems, the larger alkali metals confer dynamical stability, with NaC$_2$H$_8$ also being dynamically stable but not NaB$_2$H$_8$. This trend could be elucidated through the DFT-Chemical Pressure method~\cite{Fredrickson:2012,Berns:2014a}, which, based  on the quantum mechanical stress density, interrogates the interatomic tensions within a crystal structure that come from atomic packing. Analyzing these tensions reveals regions where shrinking or expanding the unit cell would be energetically favorable.  A particular crystal structure with DFT-relaxed coordinates represents a balance of these tensions. Here, we will investigate the role played by the alkali metals Li, Na, K, Rb, and Cs in the dynamical stability -- or lack thereof -- of the $M$B$_2$H$_8$ compounds, because a subset of them are predicted to have high $T_c$s at relatively low pressures (Figure \ref{fig:mx10tc}).

\begin{figure*}
\begin{center}
\includegraphics[width=0.85\columnwidth]{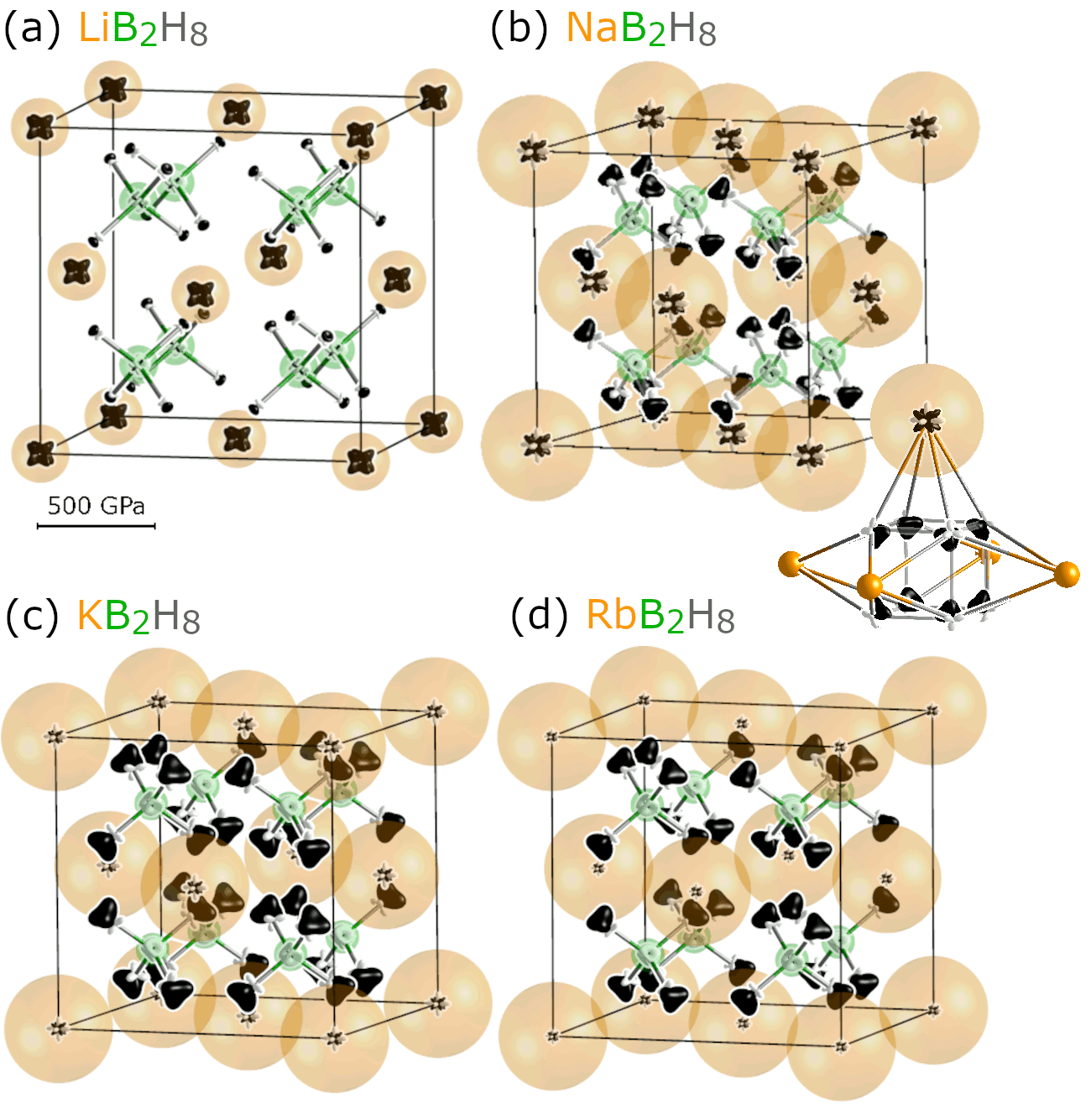}
\caption{DFT-Chemical Pressure schemes calculated for (a) LiB$_2$H$_8$, (b) NaB$_2$H$_8$, (c) KB$_2$H$_8$ and (d) RbB$_2$H$_8$ at 0~GPa. Chemical pressures surrounding each atom are projected onto spherical harmonic functions, with black lobes corresponding to negative (favoring contraction) and white lobes to positive (favoring expansion) chemical pressures. A scalebar is included for comparison; all CP schemes have been plotted to the same scale. $M$ and B atoms are highlighted with orange and green translucent spheres. As the size of the $M$ atom increases, the magnitude of the CP features decreases, while the features on the B and H atoms are relatively consistent. Positive CP occurs between the B and H atoms as well as between the $M$ and H atoms (inset to (b) shown to highlight the $M$-H interaction), with negative CP mostly within the H lattice as well as along $M$-B contacts. }
    \label{fig:CPfigure}
    \end{center}
\end{figure*}

In a DFT-Chemical Pressure scheme, the internal conflicts of atomic size in a particular crystal structure are displayed as ``chemical pressures", whose resolution can drive structural distortions or phase transitions~\cite{Berns:2014,Zurek:2022f,Lim:2023}. Here, the CP schemes in question are shown in Figure~\ref{fig:CPfigure}. In these images, each atom is surrounded by a black-and-white surface whose shape is derived from spherical harmonic functions, with the $M$ (= Li, Na, K, or Rb) and B atoms further surrounded by orange and green translucent spheres for visual clarity. The distance from the atom to the spherical-harmonic-like surface represents the magnitude of CP in that direction, with the character being represented by the color of the surface: black for negative CP, where interatomic distances are too long and shrinking of the local coordination would be favorable, and white for positive CP, where the coordination environment is overly cramped and interatomic distances are too short.

Overall, the DFT-CP schemes calculated for the four $M$B$_2$H$_8$ phases are qualitatively similar. The alkali metal atoms display positive, but small, CP lobes along the three Cartesian axes, with those for Li being particularly tiny due to the poor fit of the small cation between the BH$_4^-$ units.  These positive CPs, directed towards the nearby H atoms from surrounding BH$_4^-$ units, are echoed in the most prominent positive CP lobes on the H atoms directed right back at the $M$ atoms (Figure~\ref{fig:CPfigure}b). Negative CPs fill in the rest of the space around the metal atoms, loosely directed towards the B atoms. The CP features decorating the B atoms, which are mainly due to the close B-H interactions, are fairly consistent in magnitude across the $M$B$_2$H$_8$ series, with slight negative CP directed towards the $M$ atoms and stronger positive CP towards the H. On the H atoms, negative CPs are directed towards the H atoms in neighboring BH$_4^-$ units, while between the B and H atoms are positive CPs.

The consistency of the B-H CP features across the structural series indicates a generally rigid BH$_4^-$ subunit -- in this case, one held in place by strong covalent bonds. How well the $M$ atom fits into the interstices of the BH$_4^-$ sublattice should then govern the dynamical stability. For each member of the structural series shown, the net CP on the $M$ atoms is negative, indicating a coordination environment that is roomy rather than constrained. However, the magnitude of the CP features decorating the $M$ atoms decreases as their size increases, with the CP lobes on the K and Rb atoms being much reduced in comparison to Li and Na. This reflects an improved fit of the alkali metal atom to the interstices of the BH$_4^-$ network, borne out in a dynamically stable structure.

In fact, the DFT-CP schemes of the $M$B$_2$H$_8$ phases can provide additional information about the lattice dynamics (to be interrogated in detail below). Negative and positive CPs, especially when perpendicular to one another, are correlated with soft and hard phonon modes, respectively. High frequency modes in the $M$B$_2$H$_8$ phases examined here correspond to B-H stretches, which we expect from the CP picture to be high in energy due to the positive CP between B and H atoms. Furthermore, as discussed below, the EPC in KB$_2$H$_8$ has strong contributions from twisting motions of the H atoms in the BH$_4^-$ units. Such low-frequency motions lie along the directions of negative H-H CP -- and for LiB$_2$H$_8$ and NaB$_2$H$_8$, dip into the imaginary regime. Perhaps, here, the smaller Li and Na atoms are not able to prevent, via steric intervention, the motions of the BH$_4^-$ clusters from driving the structure into dynamical instability. 

How do the various intercalating elements affect the electronic properties of these phases? The number of electrons that can participate in the superconducting mechanism is linked to the density of states (DOS) at the Fermi level ($E_F$), which is therefore a key descriptor in predicting the $T_c$ of conventional superconductors~\cite{Shipley:2021a}. The PBE-computed DOS at $E_F$ ($g(E_F)$) at various pressures was obtained and is tabulated in Table~S7-S8. for the studied  $M$B$_2$H$_8$ and $M$C$_2$H$_8$ phases.  Compounds containing CH$_4$ or BH$_4^-$ molecular units are typically non-metallic, but combining them with electropositive elements of the appropriate valence may result in the formation of metallic systems, whose molecular vibrations are involved in superconductivity. This strategy has been explored previously via computations on Ba$_x$(CH$_4$)$_{1-x}$ ($T_c=44$~K at 90~GPa)~\cite{Jiang:2021a}, BeCH$_4$ ($T_c=30$~K at 80~GPa~\cite{Lv:2020a}), and MgCH$_4$ ($T_c=120$~K at 105~GPa)~\cite{Tian:2015a}. Hole doping insulating Ca(BH$_4$)$_2$ with K has been predicted to yield $T_c$s as high as 110~K at ambient pressures~\cite{DiCataldo:2023a}, and various molecular or extended B-H motifs metallized by electron donation from an electropositive element have been explored in the computational search for superconductivity in compounds with covalent bonds driven metallic via charge reorganization~\cite{Kokail:2017a,Gao:2023a,DiCataldo:2020a,Li:2022a}.

\begin{figure*}
\begin{center}
\includegraphics[width=\columnwidth]{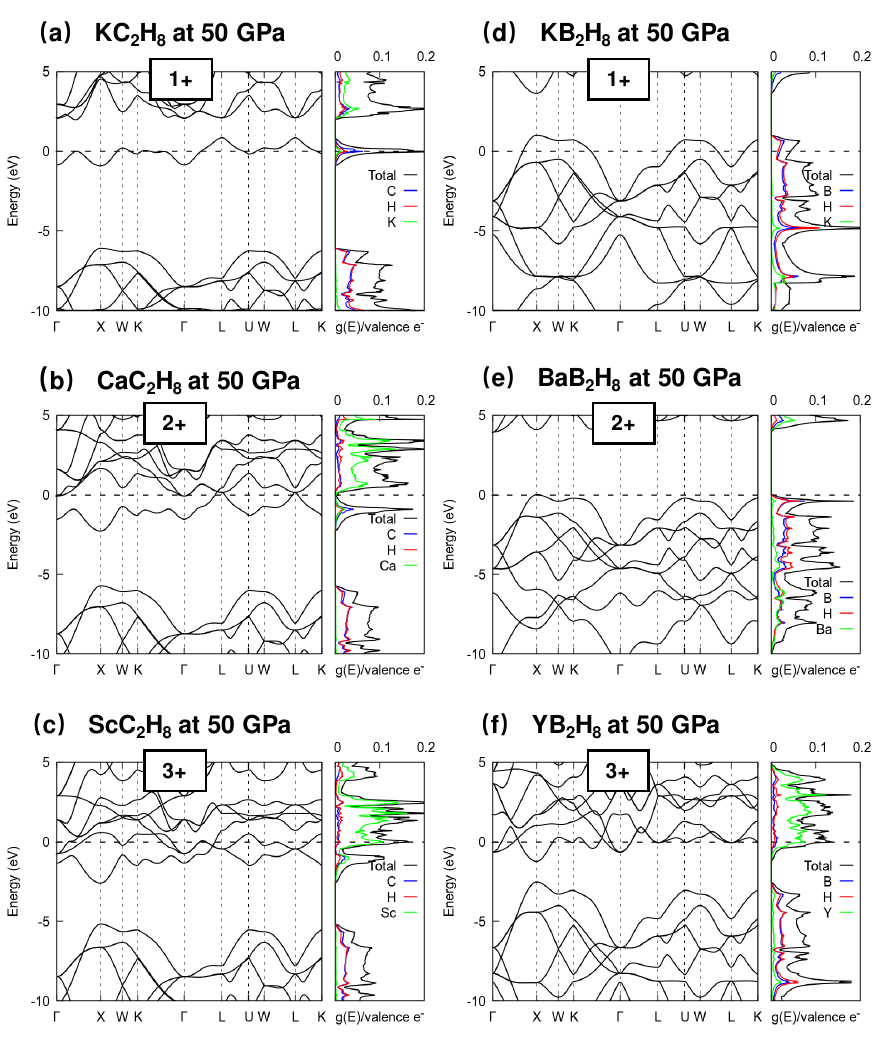}
\caption{Representative electronic band structures and projected densities of states at 50~GPa for $MX$$_2$H$_8$ phases that contain electropositive metals, $M$, with varying formal oxidation states: (a) KC$_2$H$_8$  (+1), (b) CaC$_2$H$_8$  (+2), (c) ScC$_2$H$_8$  (+3), (d) KB$_2$H$_8$  (+1), (e) BaB$_2$H$_8$  (+2), and (f) YB$_2$H$_8$  (+3). The top of the valence band (e) or Fermi energy (elsewhere) is set to 0~eV.}
    \label{fig:dos-$MX_2$H$_8$}
    \end{center}
\end{figure*}

CH$_4$ is a neutral molecule, methane. When combined with a metal intercalant, as in the $M$C$_2$H$_8$ phases, charge is transferred from the electropositive metal atoms to methane-based anti-bonding bands, rendering CH$_4$ slightly anionic.  Nevertheless, the CH$_4^{\delta-}$ units are predicted to be dynamically stable even to high pressures (ca.\ 300~GPa) in some of the studied compounds (Figure \ref{fig:mx10tc}).  
To highlight some of the noted trends, band structures and DOS plots of select structures are provided in Figure \ref{fig:dos-$MX_2$H$_8$}. In KC$_2$H$_8$ $E_F$ lies at the top of a peak in a feature of the DOS that corresponds to a half-filled band, resembling the electronic structure of $Fm\bar{3}m$ Mg$_2$IrH$_6$, which was predicted to have a $T_c$ of 160-175~K at ambient pressure~\cite{Pickard:2023f}. Full charge transfer from the potassium would result in an excess of 0.5 electrons on each methane molecule, suggesting the presence of a quarter filled doubly degenerate band crossing $E_F$. The band structure we calculate, however, is markedly different. To better understand the origin of the metallicity of this phase, we note that the lowest unoccupied molecular orbital (LUMO) of methane, of $3a_1$ symmetry, is mainly derived from an out-of-phase combination of the carbon and hydrogen $s$-orbitals. This orbital is very diffuse, Rydberg-like, with a node near the hydrogen atoms, stemming from the contribution of higher lying orbitals. 

\begin{figure*}
\begin{center}
\includegraphics[width=\columnwidth]{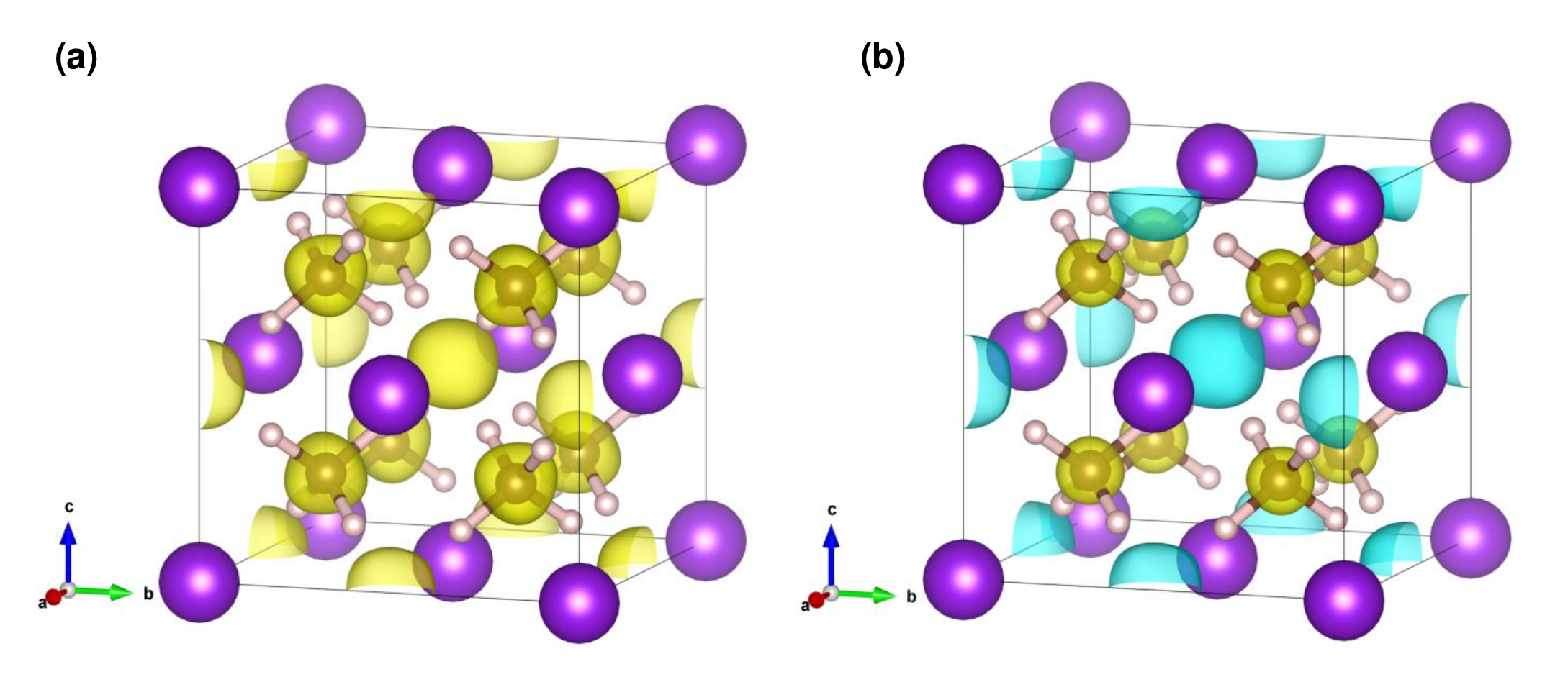}
\caption{The charge density associated with the occupied part of the metallic band in KC$_2$H$_8$ at 50~GPa. (a) Isosurface with an isovalue of 0.005~$e/\text{\AA}^3$. (b) Two-dimensional contour plot that cuts along the (1,0,1) plane colored from blue (0.000016~$e/\text{\AA}^3$) to red (0.009~$e/\text{\AA}^3$).}
    \label{fig:pcd}
    \end{center}
\end{figure*}

The LUMO of ammonia has a qualitatively similar appearance. A detailed quantum mechanical study on (NH$_3$)$_n^{\delta-}$ clusters showed that the overlap of the partially occupied LUMOs of the ammonia molecules results in a new type of bonding, denoted as H$\leftrightsquigarrow$H bonding~\cite{Zurek:2009b}. In fact, such H$\leftrightsquigarrow$H bonding, this time stemming from the overlap of the Li(NH$_3$)$_4$ singly occupied molecular orbitals (SOMOs) in lithium(0)tetraamine solid, which also possess the same Rydberg-like characteristics, leads to a build up of charge in the interstices between molecules rendering this system an electride~\cite{Zurek:2011b}.  It turns out that the peculiar electronic structure of KC$_2$H$_8$ can be explained by this same mechanism. Within it the 4b Wyckoff position is surrounded by eight hydrogen atoms from nearby methane molecules. As charge is transferred to the CH$_4$ units their diffuse LUMOs overlap forming weak, but numerous H$\leftrightsquigarrow$H interactions, leading to a build-up of charge between them. Indeed, an isosurface and contour plot of the  charge density associated with the filled portion of the metallic band illustrates that it corresponds to an interstitial blob, characteristic of an electride phase (Figure \ref{fig:pcd}).  There is one such blob per formula unit, corresponding to a single band, which is half-filled via transfer from the electropositive K atom. The H-to-blob-center distances measure 2.02~\AA{} at 4~GPa, decreasing to 1.24~\AA{} at 300~GPa. Curiously, the position of the blob also corresponds to the location of the beryllium atom in LaBeH$_8$ (and the $p$-block elements in related phases) -- in line with expectations based upon chemical template theory~\cite{templates}. CaC$_2$H$_8$ (Figure \ref{fig:dos-$MX_2$H$_8$}b) is a semi-metal resulting from the filling of this cavity-centered band. In ScC$_2$H$_8$ (Figure \ref{fig:dos-$MX_2$H$_8$}c) the Fermi level is shifted higher yet, and another set of bands crosses $E_F$, this time with substantial scandium $d$-character. 

For a particular metal atom, the electronic structure of the $M$B$_2$H$_8$ compounds is markedly different from the analogous carbon-containing systems as BH$_4^-$ is isoelectronic with CH$_4$. Indeed, KC$_2$H$_8$ is a hole doped version of potassium borohydride -- studied intensely for its potential as a hydrogen storage medium -- with a sizeable $g(E_F)$, whereas BaB$_2$H$_8$ (Figure \ref{fig:dos-$MX_2$H$_8$}e) is an insulator.  Previously experiments have identified several KBH$_4$ phases including those with cubic $Fm\overline{3}m$ (1~atm), tetragonal $P42_{1}c$ (4~GPa), and orthorhombic $Pnma$ (7~GPa) symmetry, which were reported to be semiconducting~\cite{Kumar:2008}. YB$_2$H$_8$ (Figure \ref{fig:dos-$MX_2$H$_8$}f) is isoelectronic with KC$_2$H$_8$, and its half-filled band runs-up-and-down in the same manner as the metallic band within KC$_2$H$_8$. However a key difference is that bands with Y-character, which is mostly $d$-like, hybridize with this blob-based band, and there is no gap between it and other higher lying bands. Moreover, neither YB$_2$H$_8$ nor any of the other trivalent elements considered were dynamically stable at any point in the whole pressure range studied. 

Both the carbon and boron containing compounds exhibit a high $g(E_F)$ when the metal atoms they contain are univalent or trivalent, making them potential superconductors. Compounds with bivalent elements, on the other hand, do not have an electronic structure conducive for superconductivity. For many of the methane-containing compounds increased pressure decreases $g(E_F)$ (Table~S7-S8). One exception is KC$_2$H$_8$ where the Fermi level falls right below a peak in the DOS at low pressures. As pressure increases to 100~GPa, so does $g(E_F)$, and the Fermi level becomes coincident with the peak in the DOS. At higher pressures the DOS associated with the metallic band begins to bifurcate, as a pseudogap opens up, resulting in a decreased $g(E_F)$. For KB$_2$H$_8$, RbB$_2$H$_8$ and CsB$_2$H$_8$ the DOS at $E_F$ initially decreased, then plateaued, followed by an increase to 300~GPa (for $M=$~K, Rb) or 200~GPa ($M=$~Cs). The (PBE) bandgap of  BaB$_2$H$_8$ decreased from 3.95 to 1.01~eV as pressure increased from 50-300~GPa. 

The superconducting properties of the dynamically stable, metallic phases were calculated in 50~GPa intervals and the $T_c$s were estimated using the Allen-Dynes-modified McMillan (MAD) equation~\cite{Allen:1975} and via numerical solution of the Eliashberg equations~\cite{Eliashberg:1960}, which gives a more accurate result for systems with coupling constants, $\lambda$, larger than unity.  Previous computations on YH$_6$ and YH$_9$ found almost isotropic superconducting gaps~\cite{Heil:2019a}, and therefore we have not considered anisotropy in this work. The estimated Eliashberg $T_c$s are summarized in Figure~\ref{fig:mx10tc}, where dynamically stable phases are colored according to their estimated $T_c$s. Tabulated Eliashberg and MAD $T_c$ estimates are provided in Table~S9-S11. 

Though the $M$C$_2$H$_8$ compounds containing univalent metal atoms possessed a high $g(E_F)$, their predicted $T_c$s typically fell below 20~K. This can be understood by considering the nature of their metallic bands, associated with the interstitially localized electrons (as confirmed by visualizing the charge density associated with the filled portion of the metallic band, which were analogous to the one shown in Figure \ref{fig:pcd}). Comparing to high pressure hydrides or cuprates, the $T_c$ of electrides is typically quite low~\cite{Zhangelectride,Belli:2021a}, potentially because the vibrations of the blob-based-states are not expected to yield a high EPC. The only exception is KC$_2$H$_8$ where $T_c$ is predicted to surpass 40~K above 200~GPa. For bivalent atoms, like CaC$_2$H$_8$, $E_F$ falls in a pseudogap, generally rendering them poor superconductors. The only exception to this trend is MgC$_2$H$_8$, which is estimated to have a $T_c$ of 19-35~K within its range of dynamic stability. The systems comprised of trivalent metal atoms are predicted to possess $T_c$s ranging from 40-100~K above 50~GPa, with a maximum $T_c$ of 94.0~K computed for LaC$_2$H$_8$ at 300~GPa. 

Our calculated $T_c$s for KB$_2$H$_8$ (155~K at 12~GPa), RbB$_2$H$_8$ (73~K at 50~GPa) and YC$_2$H$_8$ (58~K at 50~GPa) are comparable to previously reported values (146~K~\cite{Gao:2021}, 75~K~\cite{Li:2022s}, and 61~K~\cite{Zurek:2022f}, respectively, at the same pressures). The highest $T_c$s of 134-155~K, computed for the alkali-B$_2$H$_8$ phases, are observed for pressures below 20~GPa; at higher pressures $T_c$ falls to as low as 70~K between 20-200~GPa and then increases again to as high as 121~K at 300~GPa within the isotropic Migdal-Eliashberg theory, and neglecting quantum lattice and anharmonic effects. For the reasons described above BaB$_2$H$_8$ was not predicted to be superconducting, and combinations with other metal atoms were not dynamically stable within the harmonic approximation.

\begin{figure*}
\begin{center}
\includegraphics[width=1\columnwidth]{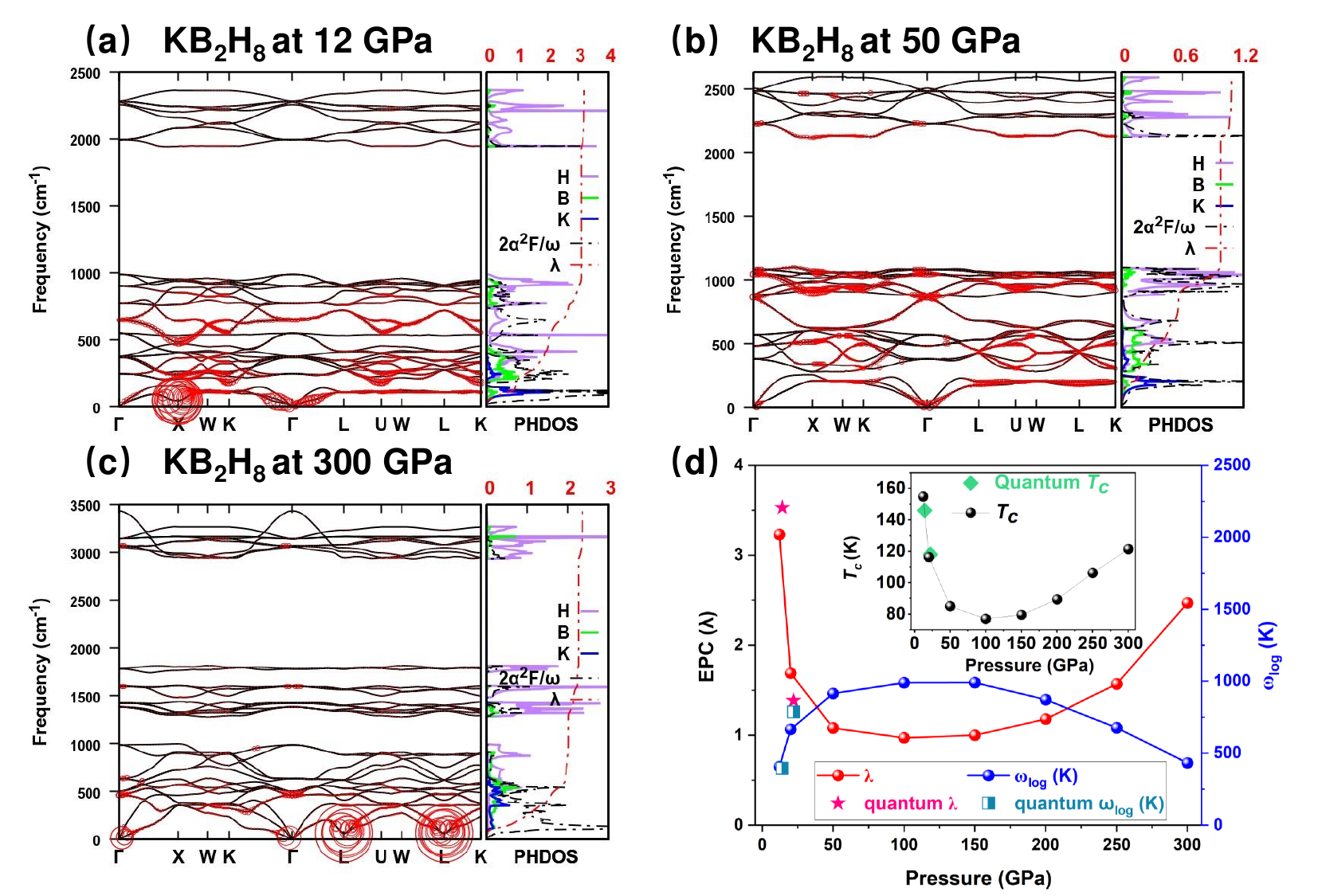}
\caption{Superconducting properties of KB$_2$H$_8$. Phonon band structures, atom projected phonon density of states (PHDOS), Eliashberg spectral function scaled by the frequency ($2\alpha^2F/\omega$), and the EPC integral ($\lambda(\omega)$) at various pressures: (a) 12~GPa, (b) 50~GPa, and (c) 300~GPa. The radius of the bubble on the phonon dispersion curve is proportional to the electron-phonon coupling constant ($\lambda_{\textbf{q}\nu}$) for the mode $\nu$ at wavevector \textbf{q}. (d) The total $\lambda$ and $\omega_\text{log}$ (logarithmic average frequency) as a function of pressure. The resulting $T_c$, as computed via the Migdal-Eliashberg formalism with $\mu^*=0.1$, is shown in the inset. Superconducting properties obtained considering quantum fluctuations at 14 and 22~GPa of quantum pressure are also provided.}
    \label{fig:epc-$MX_2$H$_8$}
    \end{center}
\end{figure*}

Previous studies of alkaline-B$_2$H$_8$ phases \cite{Gao:2021, Li:2022s, Durajski:2023first} did not look into the vibrations related to the large computed EPC (Table~S11). Herein, we examined the phonon vibrations that were key for the large $\lambda$ of KB$_2$H$_8$, as it has the highest predicted $T_c$ at the lowest pressure. Figure \ref{fig:epc-$MX_2$H$_8$} plots the phonon band structures, from low to high pressures, where red circles were utilized to represent the size of the contribution to the EPC constant at a particular wavevector $\textbf{q}$ and frequency $\nu$. At 12~GPa, a strong EPC contribution is due to the soft mode at $X$ resulting from a rocking motion of the BH$_4^{-}$ units. This soft mode remains until the pressure increases to 50~GPa, where there is no single vibration that dominates the EPC. As pressure increases to 250-300~GPa, another soft mode appears in the phonon spectrum at the $L$-point with a strong contribution to $\lambda$ involving an asymmetric scissoring within the BH$_4^{-}$ units, where one H atom bounces between two others to form short, transient, H-H contacts.

The soft phonon modes present in the band structure computed for KB$_2$H$_8$ at 12~GPa (Figure \ref{fig:epc-$MX_2$H$_8$}(a)) hint that quantum nuclear effects and the consequent anharmonicity could potentially be important in renormalizing the frequencies associated with the phonon modes and their corresponding EPC matrix elements~\cite{errea2016quantum}. Indeed, previous calculations on LaBH$_8$, whose structure, like KB$_2$H$_8$, can be derived from the parent metal $M$H$_{10}$ superhydride, is quite susceptible to these effects~\cite{belli2022impact}. Specifically, it was shown that solving the Schr\"{o}dinger equation related to the Born-Oppenheimer energy surface with the addition of the ionic kinetic energy elongates the B-H distances with a concomitant softening of the phonon modes with hydrogen character. As a result, the phase was predicted to become dynamically unstable at a notably higher pressure than predicted by phonon calculations performed under the harmonic approximation. At the same time, these effects increased the EPC and therefore the computed $T_c$. Similar conclusions have been reached for H$_3$S~\cite{errea2016quantum} and LaH$_{10}$~\cite{Errea:2020}. 

Therefore, quantum nuclear effects have been studied on the phonon band structure of KB$_{2}$H$_{8}$ at 5, 12 and 20~GPa (Figure S27). As compared to the harmonic phonons, the quantum nature of the nuclei introduces a separation between the various phonon branches with the upper branch undergoing a red shift, while the intermediate optical branch is squeezed into a smaller range of frequencies, in-line with what would be expected for strong anharmonic behavior. These variations are minimal at 20~GPa but become significant by 12~GPa. In the quantum case the instability near the $X$-point (Figure S27) is healed, and this phase is predicted to remain dynamically stable to a pressure slightly higher than
5~GPa where another instability, this time close to $\Gamma$, appears (as illustrated in the inset of Figure S27).

For classical nuclei, the electron-phonon coupling parameter, $\lambda$, logarithmic average of the phonon frequencies, $\omega_\text{log}$, and computed $T_c$ are plotted for KB$_2$H$_8$ as a function of pressure in Figure \ref{fig:epc-$MX_2$H$_8$}(d) and its inset. Results obtained for quantum nuclei are also shown (see also Table S12). Comparing with the classical results, the value of $\lambda$ is slightly increased at 12~GPa while it is reduced at 20~GPa due to the renormalizations that quantum ionic fluctuations introduce over the phonon spectra, and the length of the B-H bond. At 12~GPa quantum effects stretch the B-H bond by about 1\%. The value of $\omega_\text{log}$ is about the same at 12~GPa, but by 20~GPa quantum nuclear effects decrease it by $\sim$20\%. These variations are related to the sustained soft mode at the $X$-point and to the increased separation between the lower and central optical branches (Figure~S27). Calculating the value of $T_c$ through the Migdal-Eliashberg formalism, we find it to be slightly reduced at 12~GPa, where the variation over the phonon spectra are more significant while at 20 GPa it remains almost identical to the classical case.

\begin{figure*}
\begin{center}
\includegraphics[width=0.8\columnwidth]{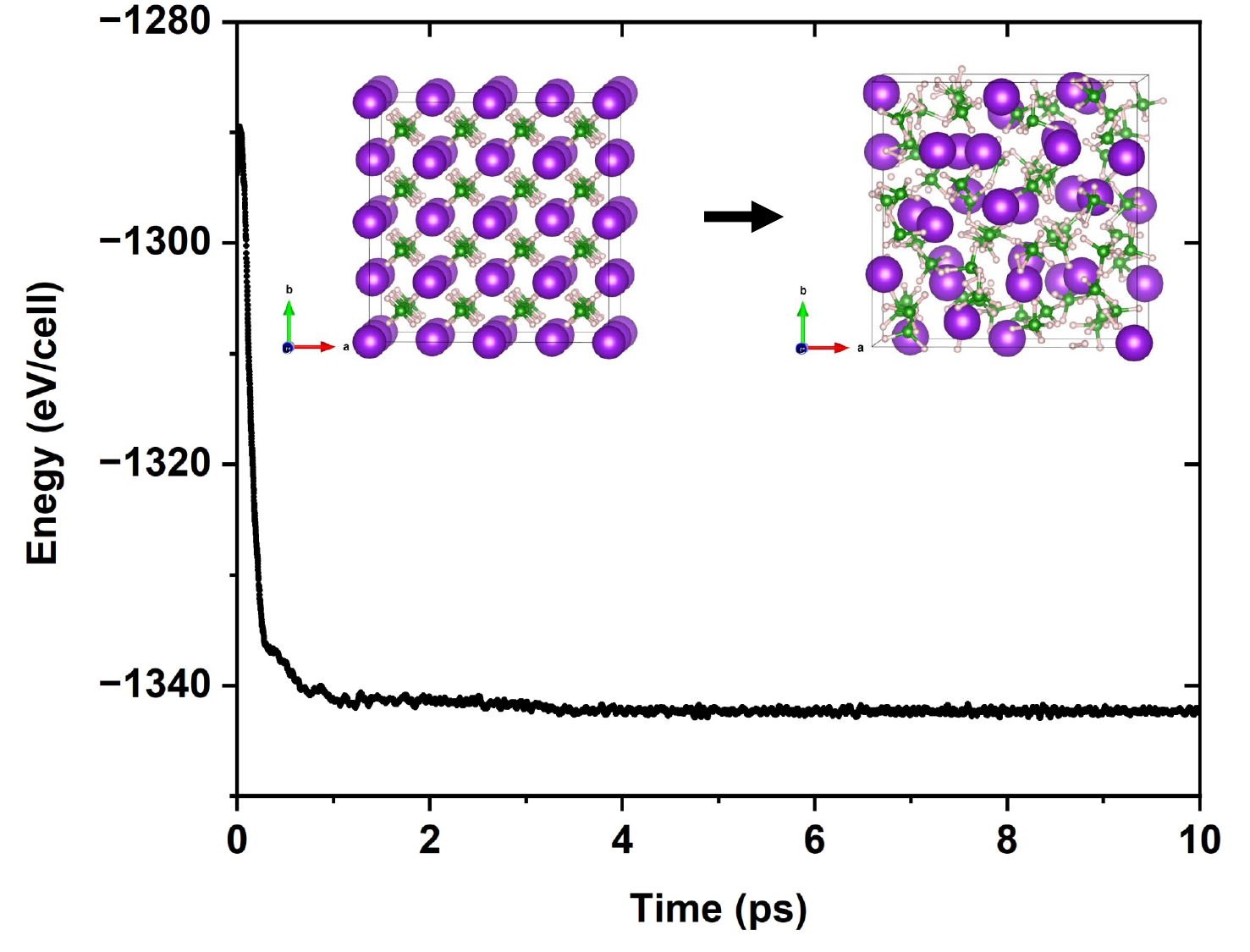}
\caption{Plot of energy versus simulation time in a first-principles molecular dynamics simulation on a supercell of KB$_{2}$H$_{8}$ at 77~K and 12~GPa. During the first picosecond of the simulation the structure overcomes a barrier and transitions to a phase containing K$^+$ and various molecular B-H containing units, as shown in the inset. K/B/H atoms are purple/green/pink.}
    \label{fig:md}
    \end{center}
\end{figure*}

Recently, a LaBeH$_8$ phase has been synthesized under pressure with a $T_c$ rivaling some of the cuprates (110~K at 80~GPa)~\cite{Song:2023s}. The KB$_{2}$H$_{8}$ phase studied here is predicted to be stable to even lower pressures and exhibit an even higher $T_c$. We therefore wondered if it could be made? Though dynamic stability is easy to assess, it is not the only important criteria for predicting the feasibility of new materials with superlative properties. Thermodynamic stability often necessitates laborious CSP searches that may, or may not, identify the thermodynamic minimum, and cannot pinpoint phases that are metastable and synthesizable. Another type of stability that is easier to probe computationally is kinetic or thermal stability as estimated via molecular dynamics (MD) simulations. Our computations showed that at 12~GPa KB$_2$H$_8$ decomposes even at 77~K (Figure \ref{fig:md}); further analysis of the structure present at the end of the simulation revealed BH$_4^-$ molecules, along with species that could be characterized as BH$_3$ units to which H$_2^{\delta -}$ was bonded side-on and (H$_3$B-H-BH$_3$)$^-$ fragments. Unsurprisingly, this decomposed phase is a PBE-semimetal. The plot of the energy evolution as a function of time of an MD run performed at 77~K illustrates a steep decrease of energy within the first picosecond, indicative of a structural transformation.  At 50~GPa, however, pressure coerced the $Fm\bar{3}m$ KB$_{2}$H$_{8}$ structure to persist up to at least 165~K, suggesting this phase, should it be made, would not fall apart below its superconducting critical temperature. Notably, while LaBeH$_8$ has a superconducting figure \cite{Pickard:2020s} of merit, $S$, of 1.236 at 80~GPa, $S=1.340$ for KB$_{2}$H$_{8}$ at 50~GPa.

\section{Conclusion}

This manuscript presents a survey of $MX_2$H$_8$ ($M$ = Li, Na, K, Rb, Cs, Be, Mg, Ca, Sr, Ba, Sc, Y, and La, $X$ = B or C) phases derived from the clathrate-like $M$H$_{10}$ superhydrides, featuring either CH$_4$ or BH$_4^-$ molecular-like subunits. Fourteen of these, six of which had not been reported before, were identified as dynamically stable at some pressure between 0-300~GPa. A DFT-Chemical pressure analysis showed that the size of the $M$ atom could play a significant role in the dynamical stability of the $M$B$_2$H$_8$ ($M=$ alkali) family.  In all cases, the electropositive metal atoms donate electrons to the CH$_4$ and BH$_4^-$ subunits, with the metal valency dictating the placement of the Fermi level and the number of occupied states. Generally, univalent or trivalent metals produce the highest density of states at the Fermi level, positioning them as the strongest candidates for superconductivity, while bivalent metals lead to insulators or semimetals. However, $M$C$_2$H$_8$ phases with univalent $M$ atoms display electride-like behavior due to the overlap of diffuse orbitals, which are unoccupied within neutral methane, that become occupied upon donation of electron density from $M$, generally suppressing their superconducting behavior due to the poor EPC associated with electrides. The highest $T_c$s are predicted to occur for the alkali-metal $M$B$_2$H$_8$ phases, with a pressure dependence showing higher $T_c$s at the extrema of the pressure range studied and decreasing at intermediate pressures. The strongest contributions to the EPC of these systems involved rocking and twisting motions of the BH$_4^-$ units. Computations on KB$_2$H$_8$ taking quantum nuclear effects into account stabilize low-frequency phonons at low pressures, bringing the onset of dynamical instability to 5, rather than 12~GPa, albeit with a slight decrease in estimated $T_c$ at 12~GPa. 

Although these results show great promise for KB$_2$H$_8$ as a warm and light superconductor, \emph{ab initio} molecular dynamics simulations reveal a crucial issue for potential synthetic attempts -- KB$_2$H$_8$ undergoes decomposition above 77~K at 12~GPa, and can only be stabilized up to 165~K with compression to 50~GPa. Nonetheless, the superconducting figure of merit predicted for KB$_2$H$_8$ rivals that of a recently synthesized LaBeH$_8$ phase~\cite{Song:2023s}. These results highlight that kinetic and thermal effects are of key importance in the prediction of viable superconductors. A vast phase-space remains to be explored to find a superconducting phase under ambient conditions, whether prompted by modification of high-pressure phases or with altogether new geometries identified by crystal structure prediction or machine learning efforts. 

\section*{Supporting Information}
Supplementary data to this article can be found online at ... . It includes the computational details, electronic band structures and densities of states, thermodynamic and thermal (molecular dynamics) stability analysis, Bader charges, structural parameters, Eliashberg spectral functions, phonon dispersion curves, EPC calculations, and details of the chemical pressure calculations.

\section*{Acknowledgments}
We acknowledge the U.S.\ National Science Foundation for financial support, specifically grants DMR-2132491 (N.G.) and DMR-2136038 (F.B.). K.H.\ is thankful to the U.S.\ Department of Energy, National Nuclear Security Administration, through the Capital-DOE Alliance Center under Cooperative Agreement DE-NA0003975 for financial support. Calculations were performed at the Center for Computational Research at SUNY Buffalo \cite{ccr}. 



%%       \citet{<label>} ==> Jones et al. [21]
%%       \citep{<label>} ==> [21]
%%

\bibliographystyle{elsarticle-num-names} 
\bibliography{mx2h8}

\begin{thebibliography}{83}
\expandafter\ifx\csname natexlab\endcsname\relax\def\natexlab#1{#1}\fi
\providecommand{\url}[1]{\texttt{#1}}
\providecommand{\href}[2]{#2}
\providecommand{\path}[1]{#1}
\providecommand{\DOIprefix}{doi:}
\providecommand{\ArXivprefix}{arXiv:}
\providecommand{\URLprefix}{URL: }
\providecommand{\Pubmedprefix}{pmid:}
\providecommand{\doi}[1]{\href{http://dx.doi.org/#1}{\path{#1}}}
\providecommand{\Pubmed}[1]{\href{pmid:#1}{\path{#1}}}
\providecommand{\bibinfo}[2]{#2}
\ifx\xfnm\relax \def\xfnm[#1]{\unskip,\space#1}\fi
%Type = Article
\bibitem[{Boeri et~al.(2022)Boeri, Hennig, Hirschfeld, Profeta, Sanna, Zurek,
  Pickett, Amsler, Dias, Eremets et~al.}]{Zurek:2020k}
\bibinfo{author}{L.~Boeri}, \bibinfo{author}{R.~Hennig},
  \bibinfo{author}{P.~Hirschfeld}, \bibinfo{author}{G.~Profeta},
  \bibinfo{author}{A.~Sanna}, \bibinfo{author}{E.~Zurek},
  \bibinfo{author}{W.~E. Pickett}, \bibinfo{author}{M.~Amsler},
  \bibinfo{author}{R.~Dias}, \bibinfo{author}{M.~I. Eremets}, et~al.,
\newblock \bibinfo{title}{The 2021 room-temperature superconductivity roadmap},
\newblock \bibinfo{journal}{J. Phys.: Condens. Matter} \bibinfo{volume}{34}
  (\bibinfo{year}{2022}) \bibinfo{pages}{183002}.
%Type = Article
\bibitem[{Ashcroft(2004)}]{Ashcroft:2004a}
\bibinfo{author}{N.~W. Ashcroft},
\newblock \bibinfo{title}{Hydrogen dominant metallic alloys: High temperature
  superconductors?},
\newblock \bibinfo{journal}{Phys. Rev. Lett.} \bibinfo{volume}{92}
  (\bibinfo{year}{2004}) \bibinfo{pages}{187002}.
%Type = Article
\bibitem[{Drozdov et~al.(2015)Drozdov, Eremets, Troyan, Ksenofontov, and
  Shylin}]{Drozdov:2015}
\bibinfo{author}{A.~P. Drozdov}, \bibinfo{author}{M.~I. Eremets},
  \bibinfo{author}{I.~A. Troyan}, \bibinfo{author}{V.~Ksenofontov},
  \bibinfo{author}{S.~I. Shylin},
\newblock \bibinfo{title}{Conventional superconductivity at 203 {Kelvin} at
  high pressures in the sulfur hydride system},
\newblock \bibinfo{journal}{Nature} \bibinfo{volume}{525}
  (\bibinfo{year}{2015}) \bibinfo{pages}{73--76}.
%Type = Article
\bibitem[{Somayazulu et~al.(2019)Somayazulu, Ahart, Mishra, Geballe, Baldini,
  Meng, Struzhkin, and Hemley}]{Somayazulu:2019}
\bibinfo{author}{M.~Somayazulu}, \bibinfo{author}{M.~Ahart},
  \bibinfo{author}{A.~K. Mishra}, \bibinfo{author}{Z.~M. Geballe},
  \bibinfo{author}{M.~Baldini}, \bibinfo{author}{Y.~Meng},
  \bibinfo{author}{V.~V. Struzhkin}, \bibinfo{author}{R.~J. Hemley},
\newblock \bibinfo{title}{Evidence for superconductivity above 260~{K} in
  lanthanum superhydride at megabar pressures},
\newblock \bibinfo{journal}{Phys. Rev. Lett.} \bibinfo{volume}{122}
  (\bibinfo{year}{2019}) \bibinfo{pages}{027001}.
%Type = Article
\bibitem[{Drozdov et~al.(2019)Drozdov, Kong, Minkov, Besedin, Kuzovnikov,
  Mozaffari, Balicas, Balakirev, Graf, Prakapenka et~al.}]{Drozdov:2019}
\bibinfo{author}{A.~P. Drozdov}, \bibinfo{author}{P.~P. Kong},
  \bibinfo{author}{V.~S. Minkov}, \bibinfo{author}{S.~P. Besedin},
  \bibinfo{author}{M.~A. Kuzovnikov}, \bibinfo{author}{S.~Mozaffari},
  \bibinfo{author}{L.~Balicas}, \bibinfo{author}{F.~F. Balakirev},
  \bibinfo{author}{D.~E. Graf}, \bibinfo{author}{V.~B. Prakapenka}, et~al.,
\newblock \bibinfo{title}{Superconductivity at 250~{K} in lanthanum hydride
  under high pressures},
\newblock \bibinfo{journal}{Nature} \bibinfo{volume}{569}
  (\bibinfo{year}{2019}) \bibinfo{pages}{528--531}.
%Type = Article
\bibitem[{Kong et~al.(2021)Kong, Minkov, Kuzovnikov, Drozdov, Besedin,
  Mozaffari, Balicas, Balakirev, Prakapenka, Chariton et~al.}]{Kong:2021a}
\bibinfo{author}{P.~Kong}, \bibinfo{author}{V.~S. Minkov},
  \bibinfo{author}{M.~A. Kuzovnikov}, \bibinfo{author}{A.~P. Drozdov},
  \bibinfo{author}{S.~P. Besedin}, \bibinfo{author}{S.~Mozaffari},
  \bibinfo{author}{L.~Balicas}, \bibinfo{author}{F.~F. Balakirev},
  \bibinfo{author}{V.~B. Prakapenka}, \bibinfo{author}{S.~Chariton}, et~al.,
\newblock \bibinfo{title}{Superconductivity up to 243~{K} in the
  yttrium--hydrogen system under high pressure},
\newblock \bibinfo{journal}{Nat. Commun.} \bibinfo{volume}{12}
  (\bibinfo{year}{2021}) \bibinfo{pages}{5075}.
%Type = Article
\bibitem[{Troyan et~al.(2021)Troyan, Semenok, Kvashnin, Sadakov, Sobolevskiy,
  Pudalov, Ivanova, Prakapenka, Greenberg, Gavriliuk et~al.}]{Troyan-YH4}
\bibinfo{author}{I.~A. Troyan}, \bibinfo{author}{D.~V. Semenok},
  \bibinfo{author}{A.~G. Kvashnin}, \bibinfo{author}{A.~V. Sadakov},
  \bibinfo{author}{O.~A. Sobolevskiy}, \bibinfo{author}{V.~M. Pudalov},
  \bibinfo{author}{A.~G. Ivanova}, \bibinfo{author}{V.~B. Prakapenka},
  \bibinfo{author}{E.~Greenberg}, \bibinfo{author}{A.~G. Gavriliuk}, et~al.,
\newblock \bibinfo{title}{Anomalous high--temperature superconductivity in
  {YH$_6$}},
\newblock \bibinfo{journal}{Adv. Mater.} \bibinfo{volume}{33}
  (\bibinfo{year}{2021}) \bibinfo{pages}{2006832}.
%Type = Article
\bibitem[{Ma et~al.(2022)Ma, Wang, Xie, Yang, Wang, Zhou, Liu, Yu, Zhao, Wang
  et~al.}]{Ma:CaH6}
\bibinfo{author}{L.~Ma}, \bibinfo{author}{K.~Wang}, \bibinfo{author}{Y.~Xie},
  \bibinfo{author}{X.~Yang}, \bibinfo{author}{Y.~Wang},
  \bibinfo{author}{M.~Zhou}, \bibinfo{author}{H.~Liu}, \bibinfo{author}{X.~Yu},
  \bibinfo{author}{Y.~Zhao}, \bibinfo{author}{H.~Wang}, et~al.,
\newblock \bibinfo{title}{High--temperature superconducting phase in clathrate
  calcium hydride {CaH$_6$} up to 215 {K} at a pressure of 172 {GPa}},
\newblock \bibinfo{journal}{Phys. Rev. Lett.} \bibinfo{volume}{128}
  (\bibinfo{year}{2022}) \bibinfo{pages}{167001}.
%Type = Article
\bibitem[{Li et~al.(2022)Li, He, Zhang, Wang, Zhang, Jia, Feng, Lu, Zhao, Zhang
  et~al.}]{Li:CaH6}
\bibinfo{author}{Z.~Li}, \bibinfo{author}{X.~He}, \bibinfo{author}{C.~Zhang},
  \bibinfo{author}{X.~Wang}, \bibinfo{author}{S.~Zhang},
  \bibinfo{author}{Y.~Jia}, \bibinfo{author}{S.~Feng}, \bibinfo{author}{K.~Lu},
  \bibinfo{author}{J.~Zhao}, \bibinfo{author}{J.~Zhang}, et~al.,
\newblock \bibinfo{title}{Superconductivity above 200~{K} discovered in
  superhydrides of calcium},
\newblock \bibinfo{journal}{Nat. Commun.} \bibinfo{volume}{13}
  (\bibinfo{year}{2022}) \bibinfo{pages}{2863}.
%Type = Article
\bibitem[{Semenok et~al.(2021)Semenok, Troyan, Ivanova, Kvashnin, Kruglov,
  Hanfland, Sadakov, Sobolevskiy, Pervakov, Lyubutin et~al.}]{Semenok:2021}
\bibinfo{author}{D.~V. Semenok}, \bibinfo{author}{I.~A. Troyan},
  \bibinfo{author}{A.~G. Ivanova}, \bibinfo{author}{A.~G. Kvashnin},
  \bibinfo{author}{I.~A. Kruglov}, \bibinfo{author}{M.~Hanfland},
  \bibinfo{author}{A.~V. Sadakov}, \bibinfo{author}{O.~A. Sobolevskiy},
  \bibinfo{author}{K.~S. Pervakov}, \bibinfo{author}{I.~S. Lyubutin}, et~al.,
\newblock \bibinfo{title}{Superconductivity at 253~{K} in lanthanum--yttrium
  ternary hydrides},
\newblock \bibinfo{journal}{Mater. Today} \bibinfo{volume}{48}
  (\bibinfo{year}{2021}) \bibinfo{pages}{18--28}.
%Type = Article
\bibitem[{Hilleke and Zurek(2022)}]{Zurek:2021k}
\bibinfo{author}{K.~P. Hilleke}, \bibinfo{author}{E.~Zurek},
\newblock \bibinfo{title}{Tuning chemical precompression: Theoretical design
  and crystal chemistry of novel hydrides in the quest for warm and light
  superconductivity at ambient pressures},
\newblock \bibinfo{journal}{J. Appl. Phys.} \bibinfo{volume}{131}
  (\bibinfo{year}{2022}) \bibinfo{pages}{070901}.
%Type = Article
\bibitem[{Zurek and Bi(2019)}]{Zurek:2018m}
\bibinfo{author}{E.~Zurek}, \bibinfo{author}{T.~Bi},
\newblock \bibinfo{title}{High--temperature superconductivity in alkaline and
  rare earth polyhydrides at high pressure: A theoretical perspective},
\newblock \bibinfo{journal}{J. Chem. Phys.} \bibinfo{volume}{150}
  (\bibinfo{year}{2019}) \bibinfo{pages}{050901}.
%Type = Article
\bibitem[{Flores-Livas et~al.(2020)Flores-Livas, Boeri, Sanna, Profeta, Arita,
  and Eremets}]{Flores-Livas:2020}
\bibinfo{author}{J.~A. Flores-Livas}, \bibinfo{author}{L.~Boeri},
  \bibinfo{author}{A.~Sanna}, \bibinfo{author}{G.~Profeta},
  \bibinfo{author}{R.~Arita}, \bibinfo{author}{M.~Eremets},
\newblock \bibinfo{title}{A perspective on conventional high--temperature
  superconductors at high pressure: Methods and materials},
\newblock \bibinfo{journal}{Phys. Rep.} \bibinfo{volume}{856}
  (\bibinfo{year}{2020}) \bibinfo{pages}{1--78}.
%Type = Article
\bibitem[{Wang et~al.(2012)Wang, Tse, Tanaka, Iitaka, and Ma}]{Wang:2012}
\bibinfo{author}{H.~Wang}, \bibinfo{author}{J.~S. Tse},
  \bibinfo{author}{K.~Tanaka}, \bibinfo{author}{T.~Iitaka},
  \bibinfo{author}{Y.~Ma},
\newblock \bibinfo{title}{Superconductive sodalite--like clathrate calcium
  hydride at high pressures},
\newblock \bibinfo{journal}{Proc. Natl. Acad. Sci. U.S.A.}
  \bibinfo{volume}{109} (\bibinfo{year}{2012}) \bibinfo{pages}{6463--6466}.
%Type = Article
\bibitem[{Liu et~al.(2017)Liu, Naumov, Hoffmann, Ashcroft, and
  Hemley}]{Liu:2017}
\bibinfo{author}{H.~Liu}, \bibinfo{author}{I.~I. Naumov},
  \bibinfo{author}{R.~Hoffmann}, \bibinfo{author}{N.~Ashcroft},
  \bibinfo{author}{R.~J. Hemley},
\newblock \bibinfo{title}{Potential high--{$T_c$} superconducting lanthanum and
  yttrium hydrides at high pressure},
\newblock \bibinfo{journal}{Natl. Acad. Sci. U. S. A.} \bibinfo{volume}{114}
  (\bibinfo{year}{2017}) \bibinfo{pages}{6990--6995}.
%Type = Article
\bibitem[{Peng et~al.(2017)Peng, Sun, Pickard, Needs, Wu, and Ma}]{Peng:2017}
\bibinfo{author}{F.~Peng}, \bibinfo{author}{Y.~Sun}, \bibinfo{author}{C.~J.
  Pickard}, \bibinfo{author}{R.~J. Needs}, \bibinfo{author}{Q.~Wu},
  \bibinfo{author}{Y.~Ma},
\newblock \bibinfo{title}{Hydrogen clathrate structures in rare earth hydrides
  at high pressures: Possible route to room-temperature superconductivity},
\newblock \bibinfo{journal}{Phys. Rev. Lett.} \bibinfo{volume}{119}
  (\bibinfo{year}{2017}) \bibinfo{pages}{107001}.
%Type = Article
\bibitem[{Duan et~al.(2014)Duan, Liu, Tian, Li, Huang, Zhao, Yu, Liu, Tian, and
  Cui}]{Duan:2014}
\bibinfo{author}{D.~Duan}, \bibinfo{author}{Y.~Liu}, \bibinfo{author}{F.~Tian},
  \bibinfo{author}{D.~Li}, \bibinfo{author}{X.~Huang},
  \bibinfo{author}{Z.~Zhao}, \bibinfo{author}{H.~Yu}, \bibinfo{author}{B.~Liu},
  \bibinfo{author}{W.~Tian}, \bibinfo{author}{T.~Cui},
\newblock \bibinfo{title}{Pressure--induced metallization of dense
  {(H$_2$S)$_2$H$_2$} with high--{$T_c$} superconductivity},
\newblock \bibinfo{journal}{Sci. Rep.} \bibinfo{volume}{4}
  (\bibinfo{year}{2014}) \bibinfo{pages}{6968}.
%Type = Incollection
\bibitem[{Bi et~al.(2019)Bi, Zarifi, Terpstra, and Zurek}]{Zurek:2018d}
\bibinfo{author}{T.~Bi}, \bibinfo{author}{N.~Zarifi},
  \bibinfo{author}{T.~Terpstra}, \bibinfo{author}{E.~Zurek},
\newblock \bibinfo{title}{The search for superconductivity in high pressure
  hydrides},
\newblock in: \bibinfo{editor}{J.~Reedijk} (Ed.), \bibinfo{booktitle}{Reference
  Module in Chemistry, Molecular Sciences and Chemical Engineering},
  \bibinfo{publisher}{Elsevier}, \bibinfo{address}{Waltham, MA},
  \bibinfo{year}{2019}, pp. \bibinfo{pages}{1--36}.
%Type = Article
\bibitem[{Salzbrenner et~al.(2023)Salzbrenner, Joo, Conway, Cooke, Zhu,
  Matraszek, Witt, and Pickard}]{Salzbrenner:2023}
\bibinfo{author}{P.~T. Salzbrenner}, \bibinfo{author}{S.~H. Joo},
  \bibinfo{author}{L.~J. Conway}, \bibinfo{author}{P.~I.~C. Cooke},
  \bibinfo{author}{B.~Zhu}, \bibinfo{author}{M.~P. Matraszek},
  \bibinfo{author}{W.~C. Witt}, \bibinfo{author}{C.~J. Pickard},
\newblock \bibinfo{title}{Developments and further applications of ephemeral
  data derived potentials},
\newblock \bibinfo{journal}{J. Chem. Phys.} \bibinfo{volume}{159}
  (\bibinfo{year}{2023}) \bibinfo{pages}{144801}.
%Type = Article
\bibitem[{Ferreira et~al.(2023)Ferreira, Conway, Cucciari, Di~Cataldo,
  Giannessi, Kogler, Eleno, Pickard, Heil, and Boeri}]{Ferreira:2023}
\bibinfo{author}{P.~P. Ferreira}, \bibinfo{author}{L.~J. Conway},
  \bibinfo{author}{A.~Cucciari}, \bibinfo{author}{S.~Di~Cataldo},
  \bibinfo{author}{F.~Giannessi}, \bibinfo{author}{E.~Kogler},
  \bibinfo{author}{L.~T.~F. Eleno}, \bibinfo{author}{C.~J. Pickard},
  \bibinfo{author}{C.~Heil}, \bibinfo{author}{L.~Boeri},
\newblock \bibinfo{title}{Search for ambient superconductivity in the {Lu-N-H}
  system},
\newblock \bibinfo{journal}{Nat. Commun.} \bibinfo{volume}{14}
  (\bibinfo{year}{2023}) \bibinfo{pages}{5367}.
%Type = Article
\bibitem[{Dolui et~al.(2023)Dolui, Conway, Heil, Strobel, Prasankumar, and
  Pickard}]{Pickard:2023f}
\bibinfo{author}{K.~Dolui}, \bibinfo{author}{L.~J. Conway},
  \bibinfo{author}{C.~Heil}, \bibinfo{author}{T.~A. Strobel},
  \bibinfo{author}{R.~Prasankumar}, \bibinfo{author}{C.~J. Pickard},
\newblock \bibinfo{title}{Feasible route to high--temperature ambient-pressure
  hydride superconductivity},
\newblock \bibinfo{journal}{arXiv preprint}  (\bibinfo{year}{2023})
  \bibinfo{pages}{arXiv:2310.07562}.
%Type = Article
\bibitem[{Yan et~al.(2020)Yan, Bi, Geng, Wang, and Zurek}]{Yan:2020}
\bibinfo{author}{Y.~Yan}, \bibinfo{author}{T.~Bi}, \bibinfo{author}{N.~Geng},
  \bibinfo{author}{X.~Wang}, \bibinfo{author}{E.~Zurek},
\newblock \bibinfo{title}{A metastable {CaSH$_3$} phase composed of {HS}
  honeycomb sheets that is superconducting under pressure},
\newblock \bibinfo{journal}{J. Phys. Chem. Lett.} \bibinfo{volume}{11}
  (\bibinfo{year}{2020}) \bibinfo{pages}{9629--9636}.
%Type = Article
\bibitem[{Geng et~al.(2022)Geng, Bi, and Zurek}]{Geng:2022}
\bibinfo{author}{N.~Geng}, \bibinfo{author}{T.~Bi}, \bibinfo{author}{E.~Zurek},
\newblock \bibinfo{title}{Structural diversity and superconductivity in
  {S--P--H} ternary hydrides under pressure},
\newblock \bibinfo{journal}{J. Phys. Chem. C} \bibinfo{volume}{126}
  (\bibinfo{year}{2022}) \bibinfo{pages}{7208--7220}.
%Type = Article
\bibitem[{Ge et~al.(2020)Ge, Zhang, Dias, Hemley, and Yao}]{Ge:2020a}
\bibinfo{author}{Y.~Ge}, \bibinfo{author}{F.~Zhang}, \bibinfo{author}{R.~P.
  Dias}, \bibinfo{author}{R.~J. Hemley}, \bibinfo{author}{Y.~Yao},
\newblock \bibinfo{title}{Hole--doped room--temperature superconductivity in
  {H$_3$S$_{1-x}$Z$_x$ (Z=C, Si)}},
\newblock \bibinfo{journal}{Mater. Today Phys.} \bibinfo{volume}{15}
  (\bibinfo{year}{2020}) \bibinfo{pages}{100330}.
%Type = Article
\bibitem[{Wei et~al.(2023)Wei, Hao, Bergara, Zurek, Liang, Wang, Song, Li,
  Wang, Gao et~al.}]{Zurek:2023a}
\bibinfo{author}{X.~Wei}, \bibinfo{author}{X.~Hao},
  \bibinfo{author}{A.~Bergara}, \bibinfo{author}{E.~Zurek},
  \bibinfo{author}{X.~Liang}, \bibinfo{author}{L.~Wang},
  \bibinfo{author}{X.~Song}, \bibinfo{author}{P.~Li},
  \bibinfo{author}{L.~Wang}, \bibinfo{author}{G.~Gao}, et~al.,
\newblock \bibinfo{title}{Designing ternary superconducting hydrides with
  {A}15-type structure at moderate pressures},
\newblock \bibinfo{journal}{Mater. Today Phys.} \bibinfo{volume}{34}
  (\bibinfo{year}{2023}) \bibinfo{pages}{101086}.
%Type = Article
\bibitem[{Lucrezi et~al.(2022)Lucrezi, Di~Cataldo, von~der Linden, Boeri, and
  Heil}]{Lucrezi:2022}
\bibinfo{author}{R.~Lucrezi}, \bibinfo{author}{S.~Di~Cataldo},
  \bibinfo{author}{W.~von~der Linden}, \bibinfo{author}{L.~Boeri},
  \bibinfo{author}{C.~Heil},
\newblock \bibinfo{title}{In-silico synthesis of lowest-pressure high--{$T_c$}
  ternary superhydrides},
\newblock \bibinfo{journal}{npj Computational Materials} \bibinfo{volume}{8}
  (\bibinfo{year}{2022}) \bibinfo{pages}{119}.
%Type = Article
\bibitem[{Zhang et~al.(2022)Zhang, Cui, Hutcheon, Shipley, Song, Du, Kresin,
  Duan, Pickard, and Yao}]{Zhang:2021}
\bibinfo{author}{Z.~Zhang}, \bibinfo{author}{T.~Cui}, \bibinfo{author}{M.~J.
  Hutcheon}, \bibinfo{author}{A.~M. Shipley}, \bibinfo{author}{H.~Song},
  \bibinfo{author}{M.~Du}, \bibinfo{author}{V.~Z. Kresin},
  \bibinfo{author}{D.~Duan}, \bibinfo{author}{C.~J. Pickard},
  \bibinfo{author}{Y.~Yao},
\newblock \bibinfo{title}{Design principles for high temperature
  superconductors with hydrogen-based alloy backbone at moderate pressure},
\newblock \bibinfo{journal}{Phys. Rev. Lett.} \bibinfo{volume}{128}
  (\bibinfo{year}{2022}) \bibinfo{pages}{047001}.
%Type = Article
\bibitem[{Liang et~al.(2021)Liang, Bergara, Wei, Song, Wang, Sun, Liu, Hemley,
  Wang, Gao, and pthers}]{Liang:2021a}
\bibinfo{author}{X.~Liang}, \bibinfo{author}{A.~Bergara},
  \bibinfo{author}{X.~Wei}, \bibinfo{author}{X.~Song},
  \bibinfo{author}{L.~Wang}, \bibinfo{author}{R.~Sun},
  \bibinfo{author}{H.~Liu}, \bibinfo{author}{R.~J. Hemley},
  \bibinfo{author}{L.~Wang}, \bibinfo{author}{G.~Gao},
  \bibinfo{author}{pthers},
\newblock \bibinfo{title}{Prediction of high--{$T_c$} superconductivity in
  ternary lanthanum borohydrides},
\newblock \bibinfo{journal}{Phys. Rev. B.} \bibinfo{volume}{104}
  (\bibinfo{year}{2021}) \bibinfo{pages}{134501}.
%Type = Article
\bibitem[{Di~Cataldo et~al.(2021)Di~Cataldo, Heil, von~der Linden, and
  Boeri}]{Di:2021bh}
\bibinfo{author}{S.~Di~Cataldo}, \bibinfo{author}{C.~Heil},
  \bibinfo{author}{W.~von~der Linden}, \bibinfo{author}{L.~Boeri},
\newblock \bibinfo{title}{{LaBH$_8$}: Towards high--{$T_c$} low-pressure
  superconductivity in ternary superhydrides},
\newblock \bibinfo{journal}{Phys. Rev. B} \bibinfo{volume}{104}
  (\bibinfo{year}{2021}) \bibinfo{pages}{L020511}.
%Type = Article
\bibitem[{Song et~al.(2023)Song, Bi, Nakamoto, Shimizu, Liu, Zou, Liu, Wang,
  and Ma}]{Song:2023s}
\bibinfo{author}{Y.~Song}, \bibinfo{author}{J.~Bi},
  \bibinfo{author}{Y.~Nakamoto}, \bibinfo{author}{K.~Shimizu},
  \bibinfo{author}{H.~Liu}, \bibinfo{author}{B.~Zou}, \bibinfo{author}{G.~Liu},
  \bibinfo{author}{H.~Wang}, \bibinfo{author}{Y.~Ma},
\newblock \bibinfo{title}{Stoichiometric ternary superhydride {LaBeH$_8$} as a
  new template for high--temperature superconductivity at 110 {K} under 80
  {GPa}},
\newblock \bibinfo{journal}{Physical Review Letters} \bibinfo{volume}{130}
  (\bibinfo{year}{2023}) \bibinfo{pages}{266001}.
%Type = Article
\bibitem[{Jiang et~al.(2022)Jiang, Hai, Tian, Ding, Feng, Yang, Chen, and
  Zhong}]{Jiang:2022}
\bibinfo{author}{M.-J. Jiang}, \bibinfo{author}{Y.-L. Hai},
  \bibinfo{author}{H.-L. Tian}, \bibinfo{author}{H.-B. Ding},
  \bibinfo{author}{Y.-J. Feng}, \bibinfo{author}{C.-L. Yang},
  \bibinfo{author}{X.-J. Chen}, \bibinfo{author}{G.-H. Zhong},
\newblock \bibinfo{title}{High--temperature superconductivity below 100 {GPa}
  in ternary {C}-based hydride {MC$_2$H$_8$} with molecular crystal
  characteristics ({M= Na, K, Mg, Al, and Ga})},
\newblock \bibinfo{journal}{Phys. Rev. B} \bibinfo{volume}{105}
  (\bibinfo{year}{2022}) \bibinfo{pages}{104511}.
%Type = Article
\bibitem[{Hilleke and Zurek(2022)}]{Zurek:2022f}
\bibinfo{author}{K.~Hilleke}, \bibinfo{author}{E.~Zurek},
\newblock \bibinfo{title}{Rational design of superconducting metal hydrides via
  chemical pressure tuning},
\newblock \bibinfo{journal}{Angew. Chem. Int. Ed.} \bibinfo{volume}{61}
  (\bibinfo{year}{2022}) \bibinfo{pages}{e202207589}.
%Type = Article
\bibitem[{Durajski and Szcz{\c e}{\'s}niak(2021)}]{Durajski:2021}
\bibinfo{author}{A.~P. Durajski}, \bibinfo{author}{R.~Szcz{\c e}{\'s}niak},
\newblock \bibinfo{title}{New superconducting superhydride {LaC$_2$H$_8$} at
  relatively low stabilization pressure},
\newblock \bibinfo{journal}{Phys. Chem. Chem. Phys.} \bibinfo{volume}{23}
  (\bibinfo{year}{2021}) \bibinfo{pages}{25070--25074}.
%Type = Article
\bibitem[{Li et~al.(2022)Li, Wang, Sun, Lu, and Peng}]{Li:2022s}
\bibinfo{author}{S.~Li}, \bibinfo{author}{H.~Wang}, \bibinfo{author}{W.~Sun},
  \bibinfo{author}{C.~Lu}, \bibinfo{author}{F.~Peng},
\newblock \bibinfo{title}{Superconductivity in compressed ternary alkaline
  boron hydrides},
\newblock \bibinfo{journal}{Phys. Rev. B} \bibinfo{volume}{105}
  (\bibinfo{year}{2022}) \bibinfo{pages}{224107}.
%Type = Article
\bibitem[{Gao et~al.(2021)Gao, Yan, Lu, and Xiang}]{Gao:2021}
\bibinfo{author}{M.~Gao}, \bibinfo{author}{X.-W. Yan}, \bibinfo{author}{Z.-Y.
  Lu}, \bibinfo{author}{T.~Xiang},
\newblock \bibinfo{title}{Phonon-mediated high--temperature superconductivity
  in the ternary borohydride {KB$_2$H$_8$} under pressure near 12 {GPa}},
\newblock \bibinfo{journal}{Phys. Rev. B} \bibinfo{volume}{104}
  (\bibinfo{year}{2021}) \bibinfo{pages}{L100504}.
%Type = Article
\bibitem[{Wan and Zhang(2022)}]{Wan:2022}
\bibinfo{author}{Z.~Wan}, \bibinfo{author}{R.~Zhang},
\newblock \bibinfo{title}{Metallization of hydrogen by intercalating ammonium
  ions in metal fcc lattices at lower pressure},
\newblock \bibinfo{journal}{Appl. Phys. Lett.} \bibinfo{volume}{121}
  (\bibinfo{year}{2022}) \bibinfo{pages}{192601}.
%Type = Article
\bibitem[{Geng et~al.(2023)Geng, Hilleke, Zhu, Wang, Strobel, and
  Zurek}]{Geng:2023}
\bibinfo{author}{N.~Geng}, \bibinfo{author}{K.~P. Hilleke},
  \bibinfo{author}{L.~Zhu}, \bibinfo{author}{X.~Wang}, \bibinfo{author}{T.~A.
  Strobel}, \bibinfo{author}{E.~Zurek},
\newblock \bibinfo{title}{Conventional high--temperature superconductivity in
  metallic, covalently bonded, binary-guest {C--B} clathrates},
\newblock \bibinfo{journal}{J. Am. Chem. Soc.} \bibinfo{volume}{145}
  (\bibinfo{year}{2023}) \bibinfo{pages}{1696--1706}.
%Type = Article
\bibitem[{Kresse and Hafner(1993)}]{Kresse:1993a}
\bibinfo{author}{G.~Kresse}, \bibinfo{author}{J.~Hafner},
\newblock \bibinfo{title}{\textit{Ab initio} molecular dynamics for liquid
  metals},
\newblock \bibinfo{journal}{Phys. Rev. B.} \bibinfo{volume}{47}
  (\bibinfo{year}{1993}) \bibinfo{pages}{558--561}.
%Type = Article
\bibitem[{Kresse and Joubert(1999)}]{Kresse:1999a}
\bibinfo{author}{G.~Kresse}, \bibinfo{author}{D.~Joubert},
\newblock \bibinfo{title}{From ultrasoft pseudopotentials to the projector
  augmented--wave method},
\newblock \bibinfo{journal}{Phys. Rev. B.} \bibinfo{volume}{59}
  (\bibinfo{year}{1999}) \bibinfo{pages}{1758--1775}.
%Type = Article
\bibitem[{Perdew et~al.(1996)Perdew, Burke, and Ernzerhof}]{Perdew:1996a}
\bibinfo{author}{J.~P. Perdew}, \bibinfo{author}{K.~Burke},
  \bibinfo{author}{M.~Ernzerhof},
\newblock \bibinfo{title}{Generalized gradient approximation made simple},
\newblock \bibinfo{journal}{Phys. Rev. Lett.} \bibinfo{volume}{77}
  (\bibinfo{year}{1996}) \bibinfo{pages}{3865--3868}.
%Type = Article
\bibitem[{Bl{\"o}chl(1994)}]{Blochl:1994a}
\bibinfo{author}{P.~E. Bl{\"o}chl},
\newblock \bibinfo{title}{Projector augmented--wave method},
\newblock \bibinfo{journal}{Phys. Rev. B} \bibinfo{volume}{50}
  (\bibinfo{year}{1994}) \bibinfo{pages}{17953--17979}.
%Type = Article
\bibitem[{Maintz et~al.(2016)Maintz, Deringer, Tchougr{\'e}eff, and
  Dronskowski}]{COHP-2016}
\bibinfo{author}{S.~Maintz}, \bibinfo{author}{V.~L. Deringer},
  \bibinfo{author}{A.~L. Tchougr{\'e}eff}, \bibinfo{author}{R.~Dronskowski},
\newblock \bibinfo{title}{Lobster: A tool to extract chemical bonding from
  plane-wave based {DFT}},
\newblock \bibinfo{journal}{J. Comput. Chem.} \bibinfo{volume}{37}
  (\bibinfo{year}{2016}) \bibinfo{pages}{1030--1035}.
%Type = Article
\bibitem[{Nos{\'e}(1984)}]{Nose:1984}
\bibinfo{author}{S.~Nos{\'e}},
\newblock \bibinfo{title}{A unified formulation of the constant temperature
  molecular dynamics methods},
\newblock \bibinfo{journal}{J. Chem. Phys.} \bibinfo{volume}{81}
  (\bibinfo{year}{1984}) \bibinfo{pages}{511--519}.
%Type = Article
\bibitem[{Shuichi(1991)}]{Shuichi:1991}
\bibinfo{author}{N.~Shuichi},
\newblock \bibinfo{title}{Constant temperature molecular dynamics methods},
\newblock \bibinfo{journal}{Prog. Theor. Phys. Supp.} \bibinfo{volume}{103}
  (\bibinfo{year}{1991}) \bibinfo{pages}{1--46}.
%Type = Article
\bibitem[{Hoover(1985)}]{Hoover:1985}
\bibinfo{author}{W.~G. Hoover},
\newblock \bibinfo{title}{Canonical dynamics: Equilibrium phase--space
  distributions},
\newblock \bibinfo{journal}{Phys. Rev. A} \bibinfo{volume}{31}
  (\bibinfo{year}{1985}) \bibinfo{pages}{1695--1697}.
%Type = Book
\bibitem[{Frenkel and Smit(2023)}]{Frenkel:2023}
\bibinfo{author}{D.~Frenkel}, \bibinfo{author}{B.~Smit},
  \bibinfo{title}{Understanding Molecular Simulation: from Algorithms to
  Applications}, \bibinfo{publisher}{Elsevier, Amsterdam, Netherlands},
  \bibinfo{year}{2023}.
%Type = Article
\bibitem[{Togo and Tanaka(2015)}]{Togo:2015}
\bibinfo{author}{A.~Togo}, \bibinfo{author}{I.~Tanaka},
\newblock \bibinfo{title}{First principles phonon calculations in materials
  science},
\newblock \bibinfo{journal}{Scr. Mater.} \bibinfo{volume}{108}
  (\bibinfo{year}{2015}) \bibinfo{pages}{1--5}.
%Type = Article
\bibitem[{Giannozzi et~al.(2009)Giannozzi, Baroni, Bonini, Calandra, Car,
  Cavazzoni, Ceresoli, Chiarotti, Cococcioni, Dabo et~al.}]{Giannozzi:2009}
\bibinfo{author}{P.~Giannozzi}, \bibinfo{author}{S.~Baroni},
  \bibinfo{author}{N.~Bonini}, \bibinfo{author}{M.~Calandra},
  \bibinfo{author}{R.~Car}, \bibinfo{author}{C.~Cavazzoni},
  \bibinfo{author}{D.~Ceresoli}, \bibinfo{author}{G.~L. Chiarotti},
  \bibinfo{author}{M.~Cococcioni}, \bibinfo{author}{I.~Dabo}, et~al.,
\newblock \bibinfo{title}{Quantum espresso: A modular and open--source software
  project for quantum simulations of materials},
\newblock \bibinfo{journal}{J. Phys.: Condens. Matter} \bibinfo{volume}{21}
  (\bibinfo{year}{2009}) \bibinfo{pages}{395502}.
%Type = Article
\bibitem[{Dal~Corso(2014)}]{DalCorso:2014}
\bibinfo{author}{A.~Dal~Corso},
\newblock \bibinfo{title}{Pseudopotentials periodic table: From {H} to {Pu}},
\newblock \bibinfo{journal}{Comput. Mater. Sci.} \bibinfo{volume}{95}
  (\bibinfo{year}{2014}) \bibinfo{pages}{337--350}.
%Type = Article
\bibitem[{Troullier and Martins(1991)}]{Troullier:1991}
\bibinfo{author}{N.~Troullier}, \bibinfo{author}{J.~L. Martins},
\newblock \bibinfo{title}{Efficient pseudopotentials for plane--wave
  calculations},
\newblock \bibinfo{journal}{Phys. Rev. B} \bibinfo{volume}{43}
  (\bibinfo{year}{1991}) \bibinfo{pages}{1993--2006}.
%Type = Article
\bibitem[{Methfessel and Paxton(1989)}]{Methfessel:1989}
\bibinfo{author}{M.~Methfessel}, \bibinfo{author}{A.~T. Paxton},
\newblock \bibinfo{title}{High--precision sampling for brillouin--zone
  integration in metals},
\newblock \bibinfo{journal}{Phys. Rev. B} \bibinfo{volume}{40}
  (\bibinfo{year}{1989}) \bibinfo{pages}{3616--3621}.
%Type = Article
\bibitem[{Allen and Dynes(1975)}]{Allen:1975}
\bibinfo{author}{P.~B. Allen}, \bibinfo{author}{R.~C. Dynes},
\newblock \bibinfo{title}{Transition temperature of strong--coupled
  superconductors reanalyzed},
\newblock \bibinfo{journal}{Phys. Rev. B} \bibinfo{volume}{12}
  (\bibinfo{year}{1975}) \bibinfo{pages}{905--922}.
%Type = Article
\bibitem[{Eliashberg(1960)}]{Eliashberg:1960}
\bibinfo{author}{G.~M. Eliashberg},
\newblock \bibinfo{title}{Interactions between electrons and lattice vibrations
  in a superconductor},
\newblock \bibinfo{journal}{Sov. Phys. JETP} \bibinfo{volume}{11}
  (\bibinfo{year}{1960}) \bibinfo{pages}{696--702}.
%Type = Article
\bibitem[{Monacelli et~al.(2021)Monacelli, Bianco, Cherubini, Calandra, Errea,
  and Mauri}]{Monacelli:2021t}
\bibinfo{author}{L.~Monacelli}, \bibinfo{author}{R.~Bianco},
  \bibinfo{author}{M.~Cherubini}, \bibinfo{author}{M.~Calandra},
  \bibinfo{author}{I.~Errea}, \bibinfo{author}{F.~Mauri},
\newblock \bibinfo{title}{The stochastic self-consistent harmonic
  approximation: Calculating vibrational properties of materials with full
  quantum and anharmonic effects},
\newblock \bibinfo{journal}{J. Phys.: Condens. Matter} \bibinfo{volume}{33}
  (\bibinfo{year}{2021}) \bibinfo{pages}{363001}.
%Type = Article
\bibitem[{Fredrickson(2012)}]{Fredrickson:2012}
\bibinfo{author}{D.~C. Fredrickson},
\newblock \bibinfo{title}{D{F}{T}-{C}hemical {P}ressure analysis: Visualizing
  the role of atomic size in shaping the structures of inorganic materials},
\newblock \bibinfo{journal}{J. Am. Chem. Soc.} \bibinfo{volume}{134}
  (\bibinfo{year}{2012}) \bibinfo{pages}{5991--5999}.
%Type = Article
\bibitem[{Berns et~al.(2014)Berns, Engelkemier, Guo, Kilduff, and
  Fredrickson}]{Berns:2014a}
\bibinfo{author}{V.~M. Berns}, \bibinfo{author}{J.~Engelkemier},
  \bibinfo{author}{Y.~Guo}, \bibinfo{author}{B.~J. Kilduff},
  \bibinfo{author}{D.~C. Fredrickson},
\newblock \bibinfo{title}{Progress in visualizing atomic size effects with
  {DFT}--chemical pressure analysis: From isolated atoms to trends in
  {AB$_{5}$} intermetallics},
\newblock \bibinfo{journal}{J. Chem. Theory Comput.} \bibinfo{volume}{10}
  (\bibinfo{year}{2014}) \bibinfo{pages}{3380--3392}.
%Type = Article
\bibitem[{Hilleke and Fredrickson(2018)}]{Hilleke:2018}
\bibinfo{author}{K.~P. Hilleke}, \bibinfo{author}{D.~C. Fredrickson},
\newblock \bibinfo{title}{Discerning chemical pressure amidst weak potentials:
  Vibrational modes and dumbbell/atom substitution in intermetallic
  aluminides},
\newblock \bibinfo{journal}{J. Phys. Chem. A} \bibinfo{volume}{122}
  (\bibinfo{year}{2018}) \bibinfo{pages}{8412--8426}.
%Type = Article
\bibitem[{Lu et~al.(2021)Lu, Van~Buskirk, Cheng, and Fredrickson}]{Lu:2021}
\bibinfo{author}{E.~Lu}, \bibinfo{author}{J.~Van~Buskirk},
  \bibinfo{author}{J.~Cheng}, \bibinfo{author}{D.~C. Fredrickson},
\newblock \bibinfo{title}{Tutorial on chemical pressure analysis: How atomic
  packing drives laves/zintl intergrowth in {K$_3$Au$_5$Tl}},
\newblock \bibinfo{journal}{Crystals} \bibinfo{volume}{11}
  (\bibinfo{year}{2021}) \bibinfo{pages}{906}.
%Type = Article
\bibitem[{Gonze et~al.(2020)Gonze, Amadon, Antonius, Arnardi, Baguet, Beuken,
  Bieder, Bottin, Bouchet, Bousquet, Brouwer et~al.}]{Gonze:2020}
\bibinfo{author}{X.~Gonze}, \bibinfo{author}{B.~Amadon},
  \bibinfo{author}{G.~Antonius}, \bibinfo{author}{F.~Arnardi},
  \bibinfo{author}{L.~Baguet}, \bibinfo{author}{J.-M. Beuken},
  \bibinfo{author}{J.~Bieder}, \bibinfo{author}{F.~Bottin},
  \bibinfo{author}{J.~Bouchet}, \bibinfo{author}{E.~Bousquet},
  \bibinfo{author}{Brouwer}, et~al.,
\newblock \bibinfo{title}{The {A}binit project: Impact, environment and recent
  developments},
\newblock \bibinfo{journal}{Comput. Phys. Commun.} \bibinfo{volume}{248}
  (\bibinfo{year}{2020}) \bibinfo{pages}{107042}.
%Type = Article
\bibitem[{Durajski and Szcz{\c e}{\'s}niak(2023)}]{Durajski:2023first}
\bibinfo{author}{A.~P. Durajski}, \bibinfo{author}{R.~Szcz{\c e}{\'s}niak},
\newblock \bibinfo{title}{First-principles estimation of low-pressure
  superconductivity in {KC$_2$H$_8$} ternary hydride},
\newblock \bibinfo{journal}{Phys. Status Solidi Rapid Res. Lett.}
  (\bibinfo{year}{2023}).
%Type = Article
\bibitem[{Berns and Fredrickson(2014)}]{Berns:2014}
\bibinfo{author}{V.~M. Berns}, \bibinfo{author}{D.~C. Fredrickson},
\newblock \bibinfo{title}{Structural plasticity: How intermetallics deform
  themselves in response to chemical pressure, and the complex structures that
  result},
\newblock \bibinfo{journal}{Inorg. Chem.} \bibinfo{volume}{53}
  (\bibinfo{year}{2014}) \bibinfo{pages}{10762--10771}.
%Type = Article
\bibitem[{Lim and Fredrickson(2023)}]{Lim:2023}
\bibinfo{author}{A.~Lim}, \bibinfo{author}{D.~C. Fredrickson},
\newblock \bibinfo{title}{Entropic control of bonding, guided by chemical
  pressure: Phase transitions and 18-n+m isomerism of {IrIn$_3$}},
\newblock \bibinfo{journal}{Inorg. Chem.} \bibinfo{volume}{62}
  (\bibinfo{year}{2023}) \bibinfo{pages}{10833--10846}.
%Type = Article
\bibitem[{Shipley et~al.(2021)Shipley, Hutcheon, Needs, and
  Pickard}]{Shipley:2021a}
\bibinfo{author}{A.~M. Shipley}, \bibinfo{author}{M.~J. Hutcheon},
  \bibinfo{author}{R.~J. Needs}, \bibinfo{author}{C.~J. Pickard},
\newblock \bibinfo{title}{High--throughput discovery of high--temperature
  conventional superconductors},
\newblock \bibinfo{journal}{Phys. Rev. B} \bibinfo{volume}{104}
  (\bibinfo{year}{2021}) \bibinfo{pages}{054501}.
%Type = Article
\bibitem[{Jiang et~al.(2021)Jiang, Tian, Hai, Lu, Tong, Wu, Li, Yang, and
  Zhong}]{Jiang:2021a}
\bibinfo{author}{M.-J. Jiang}, \bibinfo{author}{H.-L. Tian},
  \bibinfo{author}{Y.-L. Hai}, \bibinfo{author}{N.~Lu}, \bibinfo{author}{P.-F.
  Tong}, \bibinfo{author}{S.-Y. Wu}, \bibinfo{author}{W.-J. Li},
  \bibinfo{author}{C.-L. Yang}, \bibinfo{author}{G.-H. Zhong},
\newblock \bibinfo{title}{Phonon--mediated low--pressure superconductivity in
  ternary hydride {Ba-CH$_4$}},
\newblock \bibinfo{journal}{ACS Appl. Electron. Mater.} \bibinfo{volume}{3}
  (\bibinfo{year}{2021}) \bibinfo{pages}{4172--4179}.
%Type = Article
\bibitem[{Lv et~al.(2020)Lv, Zhang, Li, Hai, Lu, Li, and Zhong}]{Lv:2020a}
\bibinfo{author}{H.-Y. Lv}, \bibinfo{author}{S.-Y. Zhang},
  \bibinfo{author}{M.-H. Li}, \bibinfo{author}{Y.-L. Hai},
  \bibinfo{author}{N.~Lu}, \bibinfo{author}{W.-J. Li}, \bibinfo{author}{G.-H.
  Zhong},
\newblock \bibinfo{title}{Metallization and superconductivity in methane doped
  by beryllium at low pressure},
\newblock \bibinfo{journal}{Phys. Chem. Chem. Phys.} \bibinfo{volume}{22}
  (\bibinfo{year}{2020}) \bibinfo{pages}{1069--1077}.
%Type = Article
\bibitem[{Tian et~al.(2015)Tian, Duan, Sha, Liu, Yang, Liu, and
  Cui}]{Tian:2015a}
\bibinfo{author}{F.~Tian}, \bibinfo{author}{D.~Duan}, \bibinfo{author}{X.~Sha},
  \bibinfo{author}{Y.~Liu}, \bibinfo{author}{T.~Yang},
  \bibinfo{author}{B.~Liu}, \bibinfo{author}{T.~Cui},
\newblock \bibinfo{title}{Predicted structures and superconductivity of
  hypothetical {Mg-CH$_4$} compounds under high pressures},
\newblock \bibinfo{journal}{Mater. Res. Express} \bibinfo{volume}{2}
  (\bibinfo{year}{2015}) \bibinfo{pages}{046001}.
%Type = Article
\bibitem[{Di~Cataldo and Boeri(2023)}]{DiCataldo:2023a}
\bibinfo{author}{S.~Di~Cataldo}, \bibinfo{author}{L.~Boeri},
\newblock \bibinfo{title}{Metal borohydrides as ambient-pressure high--{$T_c$}
  superconductors},
\newblock \bibinfo{journal}{Phys. Rev. B} \bibinfo{volume}{107}
  (\bibinfo{year}{2023}) \bibinfo{pages}{L060501}.
%Type = Article
\bibitem[{Kokail et~al.(2017)Kokail, von~der Linden, and Boeri}]{Kokail:2017a}
\bibinfo{author}{C.~Kokail}, \bibinfo{author}{W.~von~der Linden},
  \bibinfo{author}{L.~Boeri},
\newblock \bibinfo{title}{Prediction of high--{$T_c$} conventional
  superconductivity in the ternary lithium borohydride system},
\newblock \bibinfo{journal}{Phys. Rev. Mater.} \bibinfo{volume}{1}
  (\bibinfo{year}{2017}) \bibinfo{pages}{074803}.
%Type = Article
\bibitem[{Gao et~al.(2023)Gao, Guo, Yang, X.-W., F., Lu, Xiang, and
  Lin}]{Gao:2023a}
\bibinfo{author}{M.~Gao}, \bibinfo{author}{P.-J. Guo}, \bibinfo{author}{H.-C.
  Yang}, \bibinfo{author}{Y.~X.-W.}, \bibinfo{author}{M.~F.},
  \bibinfo{author}{Z.-Y. Lu}, \bibinfo{author}{T.~Xiang},
  \bibinfo{author}{H.-Q. Lin},
\newblock \bibinfo{title}{Stabilizing a hydrogen rich superconductor at 1 {GPa}
  by charge transfer modulated virtual high--pressure effect},
\newblock \bibinfo{journal}{Phys. Rev. B} \bibinfo{volume}{107}
  (\bibinfo{year}{2023}) \bibinfo{pages}{L180501}.
%Type = Article
\bibitem[{Di~Cataldo et~al.(2020)Di~Cataldo, von~der Linden, and
  Boeri}]{DiCataldo:2020a}
\bibinfo{author}{S.~Di~Cataldo}, \bibinfo{author}{W.~von~der Linden},
  \bibinfo{author}{L.~Boeri},
\newblock \bibinfo{title}{Phase diagram and superconductivity of calcium
  borohydrides at extreme pressures},
\newblock \bibinfo{journal}{Phys. Rev. B} \bibinfo{volume}{102}
  (\bibinfo{year}{2020}) \bibinfo{pages}{014516}.
%Type = Article
\bibitem[{Li et~al.(2022)Li, Zhang, Bergara, Liu, and Yang}]{Li:2022a}
\bibinfo{author}{X.~Li}, \bibinfo{author}{X.~Zhang},
  \bibinfo{author}{A.~Bergara}, \bibinfo{author}{Y.~Liu},
  \bibinfo{author}{G.~Yang},
\newblock \bibinfo{title}{Structural and electronic properties of {Na-B-H}
  compounds at high pressure},
\newblock \bibinfo{journal}{Phys. Rev. B} \bibinfo{volume}{106}
  (\bibinfo{year}{2022}) \bibinfo{pages}{174104}.
%Type = Article
\bibitem[{Zurek et~al.(2009)Zurek, Edwards, and Hoffmann}]{Zurek:2009b}
\bibinfo{author}{E.~Zurek}, \bibinfo{author}{P.~P. Edwards},
  \bibinfo{author}{R.~Hoffmann},
\newblock \bibinfo{title}{A molecular perspective on lithium-ammonia
  solutions},
\newblock \bibinfo{journal}{Angew. Chem. Int. Ed.} \bibinfo{volume}{48}
  (\bibinfo{year}{2009}) \bibinfo{pages}{8198--8232}.
%Type = Article
\bibitem[{Zurek et~al.(2011)Zurek, Wen, and Hoffmann}]{Zurek:2011b}
\bibinfo{author}{E.~Zurek}, \bibinfo{author}{X.-D. Wen},
  \bibinfo{author}{R.~Hoffmann},
\newblock \bibinfo{title}{({B}arely) {S}olid {Li(NH$_3$)$_4$}: The electronics
  of an expanded metal},
\newblock \bibinfo{journal}{J. Am. Chem. Soc.} \bibinfo{volume}{133}
  (\bibinfo{year}{2011}) \bibinfo{pages}{3535--3547}.
%Type = Article
\bibitem[{Sun and Miao(2023)}]{templates}
\bibinfo{author}{Y.~Sun}, \bibinfo{author}{M.~Miao},
\newblock \bibinfo{title}{Chemical templates that assemble the metal
  superhydrides},
\newblock \bibinfo{journal}{Chem} \bibinfo{volume}{9} (\bibinfo{year}{2023})
  \bibinfo{pages}{443--459}.
%Type = Article
\bibitem[{Kumar et~al.(2008)Kumar, Kim, and Cornelius}]{Kumar:2008}
\bibinfo{author}{R.~S. Kumar}, \bibinfo{author}{E.~Kim}, \bibinfo{author}{A.~L.
  Cornelius},
\newblock \bibinfo{title}{Structural phase transitions in the potential
  hydrogen storage compound {KBH$_4$} under compression},
\newblock \bibinfo{journal}{J. Phys. Chem. C} \bibinfo{volume}{112}
  (\bibinfo{year}{2008}) \bibinfo{pages}{8452--8457}.
%Type = Article
\bibitem[{Heil et~al.(2019)Heil, di~Cataldo, Bachelet, and Boeri}]{Heil:2019a}
\bibinfo{author}{C.~Heil}, \bibinfo{author}{S.~di~Cataldo},
  \bibinfo{author}{G.~B. Bachelet}, \bibinfo{author}{L.~Boeri},
\newblock \bibinfo{title}{Superconductivity in sodalite-like yttrium hydride
  clathrates},
\newblock \bibinfo{journal}{Phys. Rev. B} \bibinfo{volume}{99}
  (\bibinfo{year}{2019}) \bibinfo{pages}{220502}.
%Type = Article
\bibitem[{Zhang and Yang(2020)}]{Zhangelectride}
\bibinfo{author}{X.~Zhang}, \bibinfo{author}{G.~Yang},
\newblock \bibinfo{title}{Recent advances and applications of inorganic
  electrides},
\newblock \bibinfo{journal}{J. Phys. Chem. Lett.} \bibinfo{volume}{11}
  (\bibinfo{year}{2020}) \bibinfo{pages}{3841--3852}.
%Type = Article
\bibitem[{Belli et~al.(2021)Belli, Novoa, Contreras-Garcia, and
  Errea}]{Belli:2021a}
\bibinfo{author}{F.~Belli}, \bibinfo{author}{T.~Novoa},
  \bibinfo{author}{J.~Contreras-Garcia}, \bibinfo{author}{I.~Errea},
\newblock \bibinfo{title}{Strong correlation between electronic bonding network
  and critical temperature in hydrogen--based superconductors},
\newblock \bibinfo{journal}{Nat. Commun.} \bibinfo{volume}{12}
  (\bibinfo{year}{2021}) \bibinfo{pages}{5381}.
%Type = Article
\bibitem[{Errea et~al.(2016)Errea, Calandra, Pickard, Nelson, Needs, Li, Liu,
  Zhang, Ma, and Mauri}]{errea2016quantum}
\bibinfo{author}{I.~Errea}, \bibinfo{author}{M.~Calandra},
  \bibinfo{author}{C.~J. Pickard}, \bibinfo{author}{J.~R. Nelson},
  \bibinfo{author}{R.~J. Needs}, \bibinfo{author}{Y.~Li},
  \bibinfo{author}{H.~Liu}, \bibinfo{author}{Y.~Zhang},
  \bibinfo{author}{Y.~Ma}, \bibinfo{author}{F.~Mauri},
\newblock \bibinfo{title}{Quantum hydrogen-bond symmetrization in the
  superconducting hydrogen sulfide system},
\newblock \bibinfo{journal}{Nature} \bibinfo{volume}{532}
  (\bibinfo{year}{2016}) \bibinfo{pages}{81--84}.
%Type = Article
\bibitem[{Belli and Errea(2022)}]{belli2022impact}
\bibinfo{author}{F.~Belli}, \bibinfo{author}{I.~Errea},
\newblock \bibinfo{title}{Impact of ionic quantum fluctuations on the
  thermodynamic stability and superconductivity of {LaBH$_8$}},
\newblock \bibinfo{journal}{Phys. Rev. B} \bibinfo{volume}{106}
  (\bibinfo{year}{2022}) \bibinfo{pages}{134509}.
%Type = Article
\bibitem[{Errea et~al.(2020)Errea, Belli, Monacelli, Sanna, Koretsune, Tadano,
  Bianco, Calandra, Arita, Mauri et~al.}]{Errea:2020}
\bibinfo{author}{I.~Errea}, \bibinfo{author}{F.~Belli},
  \bibinfo{author}{L.~Monacelli}, \bibinfo{author}{A.~Sanna},
  \bibinfo{author}{T.~Koretsune}, \bibinfo{author}{T.~Tadano},
  \bibinfo{author}{R.~Bianco}, \bibinfo{author}{M.~Calandra},
  \bibinfo{author}{R.~Arita}, \bibinfo{author}{F.~Mauri}, et~al.,
\newblock \bibinfo{title}{Quantum crystal structure in the 250-{K}elvin
  superconducting lanthanum hydride},
\newblock \bibinfo{journal}{Nature} \bibinfo{volume}{578}
  (\bibinfo{year}{2020}) \bibinfo{pages}{66--69}.
%Type = Article
\bibitem[{Pickard et~al.(2020)Pickard, Errea, and Eremets}]{Pickard:2020s}
\bibinfo{author}{C.~J. Pickard}, \bibinfo{author}{I.~Errea},
  \bibinfo{author}{M.~I. Eremets},
\newblock \bibinfo{title}{Superconducting hydrides under pressure},
\newblock \bibinfo{journal}{Annu. Rev. Condens. Matter Phys.}
  \bibinfo{volume}{11} (\bibinfo{year}{2020}) \bibinfo{pages}{57--76}.
%Type = Misc
\bibitem[{\text{Center for Computational Research, University at
  Buffalo.}(2023)}]{ccr}
\bibinfo{author}{\text{Center for Computational Research, University at
  Buffalo.}}, \bibinfo{year}{2023}.
  \bibinfo{note}{Http://hdl.handle.net/10477/79221 (accessed December 22nd,
  2023)}.

\end{thebibliography}


\end{document}